\documentstyle[12pt,epsf]{article}
\renewcommand{\thefootnote}{\fnsymbol{footnote}}
\newcommand{\EQ}{\begin{equation}}
\newcommand{\EN}{\end{equation}}
\newcommand{\bea}{\begin{eqnarray}}
\newcommand{\ena}{\end{eqnarray}}
\newcommand{\bear}{\begin{array}}
\newcommand{\enar}{\end{array}}

\newcommand{\vs}[1]{\vspace{#1}}

\newcommand{\lw}[1]{\smash{\lower1.5ex\hbox{#1}}}

\newcommand{\uda}{\nearrow \kern-1em \searrow}

\newcommand{\NP}[1]{Nucl.\ Phys.\ {\bf #1}}
\newcommand{\PL}[1]{Phys.\ Lett.\ {\bf #1}}

\newcommand{\PR}[1]{Phys.\ Rev.\ {\bf #1}}
\newcommand{\PRL}[1]{Phys.\ Rev.\ Lett.\ {\bf #1}}

\begin{document}
\topmargin 0pt
\oddsidemargin 5mm
\evensidemargin 5mm
\begin{titlepage}
\setcounter{page}{0}
\begin{flushright}
OU-HET 301\\
July 1998
\end{flushright}
\vs{3mm}
\begin{center}
{\Large Unified Explanation of Quark and Lepton Masses and Mixings in the 
Supersymmetric SO(10) Model}
\\
\vs{5mm}
{\bf {\sc Kin-ya Oda,\footnote{e-mail: oda@phys.wani.osaka-u.ac.jp}} 
{\sc Eiichi Takasugi,\footnote{e-mail: 
takasugi@phys.wani.osaka-u.ac.jp}} 
 {\sc Minoru Tanaka,\footnote{e-mail: 
 tanaka@phys.wani.osaka-u.ac.jp}}\\ 
 and \\
 {\sc Masaki Yoshimura\footnote{e-mail: 
 masaki@phys.wani.osaka-u.ac.jp}}
\\
\vs{4mm} 
{\em Department of Physics,\\
Osaka University\\
Toyonaka, Osaka 560 Japan}}\\
\end{center}
\vs{6mm}

\begin{abstract} 
We discussed neutrino masses and mixings in SUSY SO(10) model 
where quarks and leptons have Yukawa couplings to at least two 10 
and one $\overline{126}$ Higgs scalars. In this model, 
the Dirac and the right-handed Majorana mass terms are expressed 
by linear combinations of quark and charged lepton mass matrices, which 
then determine the neutrino mass matrix  by the see-saw mechanism. We 
show that there are various solutions to reproduce a large mixing angle 
for $\nu_\mu-\nu_\tau$ and a small mixing angle for  $\nu_e-\nu_\mu$, 
as well as the hierarchical mass spectrum of neutrinos.  
\end{abstract}
\end{titlepage}

\renewcommand{\thefootnote}{\arabic{footnote}}
\setcounter{footnote}{0}
\newpage
\section{Introduction}
Super-Kamiokande group announced the remarkable report[1], 
the evidence of the neutrino oscillation and the neutrino 
masses based on the atmospheric neutrino observation. 
The $\nu_\mu-\nu_e$ oscillation scenario is excluded by the 
CHOOZ data[2] and also the Super-Kamiokande data[1], and  
the  $\nu_\mu-\nu_\tau$ oscillation is favored, although 
the interpretation of the $\nu_\mu-\nu_{sterile}$  is possible. 
Within three neutrino scenario, they showed that 
$\sin^2 2\theta_{\nu\tau}>0.8$ and $\Delta m_{23}^2$ is  
in the range $10^{-3}{\rm eV}^2\sim 10^{-2}{\rm eV}^2$. 
On the other hand,  from the observation of the day and night 
difference of the solar neutrino flux, it seems that the small 
mixing for $\nu_e-\nu_\mu$ is favored for the solar neutrino 
problem[3]. 

In this paper, we consider these facts seriously and 
seek the scenario to reproduce the pattern of the neutrino mixings and 
the neutrino mass spectrum. In particular, we aim at how the large mixing 
between $\nu_\mu$ and $\nu_\tau$ arises, while one keeps the small 
mixing between $\nu_e$ and $\nu_\mu$ in the framework of 
SUSY SO(10) GUT models. We consider the model where fermions have 
Yukawa couplings to at least two 10 and one $\overline{126}$ Higgs scalars.  
In this scenario, the Dirac and 
the right-handed Majorana mass terms are expressed 
by linear combinations of quark and charged lepton mass matrices and 
thus   the neutrino mass matrix arising from the see-saw mechanism is 
also determined by quark and charged lepton mass matrices. In the 
basis where the u-type quark mass matrix is diagonal, the d-type 
and also the l-type(charged lepton) mass matrices are expected to be 
almost diagonal so that it is a nontrivial problem how to obtain the 
non-hierarchical neutrino mass matrix for the part 
related to the second and the third generations 
by using these hierarchical mass matrices, which is needed to 
obtain the large mixing between $\nu_\mu$ and $\nu_\tau$.  

The model which we consider have been discussed intensively 
to get  the unified description of quark and lepton masses 
and mixings.  Babu and Mohapatra[4], and Lee and Mohapatra[5] 
considered the minimal (SUSY) SO(10) GUT model, where quarks and 
leptons have 
Yukawa couplings to only one 10 and one $\overline{126}$ Higgs scalars, 
in order to get  predictions of  neutrino masses and 
mixings. Along this line, 
the texture zero analysis based on models of one 10 and 
two or three $\overline{126}$ Higgs scalars was made by Babu and Shafi[6], 
and Achiman and Greiner[7]. All these models have predicted small mixing 
angle for $\nu_\mu -\nu_\tau$. The particular interest of these 
models lies in the fact that neutrino mixings and ratios of neutrino 
masses are predicted. 

Recently, Brahmachari and Mohapatra[8] discussed that 
minimal  SUSY SO(10) models with one 10 and one $\overline{126}$ is 
unable to predict a large mixing angle for $\nu_\mu -\nu_\tau$. 
Therefore, they considered SUSY SO(10) models 
where Higgs multiplets are in two 10 and one $\overline{126}$ 
representations. 
By considering 
type II see-saw mechanism where neutrino mass matrix consists 
of the left-handed Majorana mass term and the see-saw term, they 
found a solution which predicts a large mixing angle for 
$\nu_\mu -\nu_\tau$, and a small mixing angle for $\nu_e -\nu_\mu$. 

Motivated by the work in Ref.8, we consider SUSY SO(10) models  
where quarks and leptons have Yukawa couplings to more than two 10 
and one $\overline{126}$ Higgs scalars. The model is essentially the 
same as those by Brahmachari and Mohapatra except that we do not 
consider the left-handed Majorana mass term.  We assume that 
the low energy theory of these models is the MSSM with two Higgs 
doublets which are linear combinations of the doublets in the 
10's and the $\overline{126}$. We assume that 
$H_u=\alpha_1 H_u(10_1)+\alpha_2 H_u(10_2)+\alpha_3 H_u(\overline{126})$ 
and $H_d=\beta_1 H_d(10_1)+\beta_2 H_d(10_2)+\beta_3 H_d(\overline{126})$.  
Quark and lepton masses come from the Yukawa couplings, 
\bea
W_Y=h_{i,ab}\psi_a\psi_b H_i(10) 
+f_{ab}\psi_a\psi_b \bar\Delta(\bar{126})\;,
\ena
where $\psi_a$ is the 16 dimensional fermions with the family index $a$. 
The matrices $h_j$'s and   $f$ are 3 by 3 complex symmetric matrices.  
Quark and lepton mass matrices are given by 
\bea
&M_u = \sum_j^nv_{uj} h_j + \kappa_uf\;,& \quad 
 M_d = \sum_j^nv_{dj} h_j  + \kappa_d f\;,\nonumber\\
&M_{\nu D} = \sum_j^n v_{uj}h_j -3\kappa_uf\;,& \quad 
 M_l       = \sum_j^n v_{dj} h_j  -3 \kappa_d f\;,\nonumber\\
&M_{\nu R} = v_R f\;,& 
\ena
where  $v_{u1}$ is the vacuum expectation value of $H_u(10_1)$ 
multiplied by the ratio of $H_u$ in $H_u(10_1)$. Others are 
defined similarly. They satisfy 
\bea
\sqrt{\sum_j^n (v_{uj}^2+v_{dj}^2)+\kappa_u^2+\kappa_d^2}=246{\rm GeV}\;.
\ena
The parameter $v_R$ is the scale of the right-handed Majorana neutrino 
masses. 

From Eq.(2), we obtain the relation for $n \ge 2$
\bea
M_{\nu D} &=& M_u - r (M_d-M_l)\;,\nonumber
\\ 
M_{\nu R} &=& R (M_d-M_l) \;,
\ena
where $r={\kappa_u}/{\kappa_d}$ and $ R={v_R}/{4\kappa_d}$. 
Since we consider that the neutrino mass matrix $m_\nu$ is derived by 
the see-saw mechanism as $m_\nu=-M_{\nu D}^T M_{\nu R}^{-1}M_{\nu D}$, 
$m_\nu$ is essentially determined by quark and charged lepton mass 
matrices. Only other parameters are the ratios of vacuum expectation 
values, $r$ and $R$. The parameter $R$ is used to control the 
overall normalization of neutrino masses and therefore $r$ is the 
only adjusting parameter to fix neutrino mixings and the ratios of 
neutrino masses. The models presented above are quite 
tight so that there needs some mechanism which naturally  leads to 
the large mixing between $\nu_\mu$ and $\nu_\tau$. 

The topics in this paper are as follows:

\noindent 
 (i) We show that the minimal model consisting of 
one 10 and one $\overline{126}$ Higgs is excluded. This is 
because of the inability to reproduce experimentally observed 
pattern of mass spectrum of d-type quarks and charged leptons 
simultaneously. Our reason is severer 
than the reason raised by Brahmachari and Mohapatra[8]. 
They argued that  the model is unable to predict  a large mixing angle 
between $\nu_\mu-\nu_\tau$ and thus the model is rejected. 
The outline of our discussion is given in 
Appendix A.

\noindent
 (ii) We consider the (type I) 
see-saw mechanism to obtain the neutrino mass matrix in contrast to 
Brahmachari and Mohapatra[8] who used the type II see-saw mechanism 
where the left-handed Majorana mass term is added to the see-saw 
term. We simply  avoid to introduce an extra freedom due to 
 the left-handed Majorana mass term. 

\noindent
(iii) We give a qualitative argument how to get the 
less hierarchical structure of neutrino mass matrix which 
is derived through the see-saw mechanism by using the 
Dirac and the right-handed Majorana mass terms which  are 
given as linear combinations of hierarchical u-type, d-type and 
the charged lepton mass terms.  

\noindent
(iv) We found many possible ranges of parameter $r$ to lead 
a small mixing for $\nu_e-\nu_\mu$ and a large mixing for 
$\nu_\mu-\nu_\tau$ as well as the hierarchical neutrino mass spectrum.  
Each region corresponds to slightly different 
mixing angles and neutrino masses. 

The paper is organazed as follows: In Sec.2, we explain the 
general structure of the Dirac mass matrix and the Majorana 
mass matrix, and then we explaine the strategy how to obtain 
the large mixing between $\nu_\mu$ and $\nu_\tau$. In Sec.3, 
we present the numerical analysis on whether models can 
reproduce the atomospheric neutrino as well as the solar neutrino 
experimental data. The summary is given in Sec.4.
 
\section{The unified description of quark and lepton masses and 
mixings}

In this section, we present the qualitative argument how to 
derive the large mixing between $\nu_\mu$ and $\nu_\tau$.

\subsection{ The basis of mass matrices} 

There is a freedom to choose the basis of mass matrices. 
Because neutrino mass matrices are expressed as linear 
combinations of quark and charged lepton mass matrices, we can 
 transform all mass matrices simultaneously by a unitary 
matrix such that  $M_k'\equiv U^TM_kU$ where $k=u, d, 
 l, \nu D, \nu R$ by a unitary matrix $U$. Since all 
mass matrices are symmetric ones, we choose $U$ such that it diagonalizes  
$M_u$ 
\bea
 M_u'\equiv U^TM_uU=D_u\;,
\ena
where $D_u={\rm diag}(|m_u|,|m_c|,|m_t|)$.
Next, we introduce a unitary matrix $D$ which diagonalizes $M_d$, i.e., 
$D^TM_dD=D_d$ with $D_d={\rm diag}(|m_d|,|m_s|,|m_b|)$. By using $U$ and 
$D$, the general form of the CKM matrix is given by 
$U^{\dagger}D=\phi_u^\dagger 
K\phi_d$ where $K$ is a special form of the CKM matrix and $\phi_i$ 
(i=u,d) are diagonal phase matrices. By using $K$ and phase matrices   
$M_d'$ is expressed by
\bea
 M_d'=U^TM_dU=U^TD^*D_dD^{\dagger}U=
 \phi_u^T K^* \phi_d^* D_d \phi_d^* K^\dagger \phi_u\;.
\ena
Finally, we transform all matrices by the phase matrix $\phi_u$ 
as $ \phi_u^*M_k'\phi_u^{\dagger}\equiv\tilde M_k$. 
This is  the basis which we use in below:
\bea
\tilde M_u=\tilde D_u\;, \;\tilde M_d=K^* \tilde D_d K^\dagger\;
\ena
and $\tilde M_l$, $\tilde M_{\nu D}$ and $\tilde M_{\nu R}$, 
where $\tilde D_u= \phi_u^* D_u \phi_u^{\dagger}\equiv {\rm diag}
(m_u,m_c,m_t) $  and 
$\tilde D_d= \phi_d^* D_d \phi_d^{\dagger}\equiv {\rm diag}
(m_d,m_s,m_b) $. In this basis, all quark masses are  
complex quantities. Now we have
\bea
\tilde M_{\nu D} = \tilde M_u - r \tilde M\;,\qquad
\tilde M_{\nu R} = R \tilde M \;,
\ena
where
\bea
\tilde M=\tilde M_d -\tilde M_l\;.
\ena

The left-handed neutrino mass matrix $m_\nu$ is given by 
the see-saw mechanism as
\bea
m_\nu=-\tilde M_{\nu D}^T(\tilde M_{\nu R})^{-1}\tilde M_{\nu D}\;.
\ena
Thus, neutrino mass matrix is essentially 
determined once quark and charged lepton mass matrices are given. 
Only other parameters in the model are $r$ and $R$. The parameter $R$ 
determines the overall scale of neutrino masses. Thus, the 
 parameter $r$ is the only adjusting parameter  to reproduce 
 the desired neutrino mixing angles and  neutrino mass ratios. 

\subsection{Quark and charged lepton masses and CKM mixings at the GUT 
scale}

We use the following quark and charged lepton masses, and the 
CKM mixing angles at GUT scale ($2\times 10^{16}$GeV) 
which were estimated in the minimal supersymmetric standard model with 
$\tan \beta=10$ following Fusaoka and Koide[9]:

\noindent
\bea
\matrix{|m_u|=0.00104 &|m_c|=0.302 &|m_t|=129 \cr
        |m_d|=0.00133&|m_s|= 0.0265&|m_b|= 1.00\cr
        |m_e|=0.000325&|m_\mu|=0.0686 &|m_\tau|=1.171\cr}\;,
\ena
\bea
\sin \theta_{12}=-0.2205\;,\sin \theta_{13}=0.0026\;,
\sin \theta_{23}=0.0318\;.
\ena
where fermion masses are defined in unit of GeV. 
Although these values of parameters should have errors, we neglect 
errors since our purpose is to answer whether the parameter 
ranges of $r$ exist to reproduced the desired neutrino mixings 
and neutrino spectrum. The above values of CKM mixings at GUT 
scale are given by taking into account of one loop 
contribution by keeping only $m_t$ and $m_b$. 

In the numerical analysis, we use values in Eqs.(11) and (12). 
For the qualitative analysis, we use  the Wolfenstein form 
for the CKM matrix $K$ at the GUT scale
\bea
K=\left(\matrix{
1-\frac12 \lambda^2-\frac18 \lambda^4&\lambda&
                  \Lambda \lambda^4e^{-i\delta}\cr
   -\lambda+\frac12 A^2\lambda^5&1-\frac12 \lambda^2-
              (\frac18+\frac12 A^2) \lambda^4& A\lambda^2\cr
   A\lambda^3-\Lambda \lambda^4e^{i\delta}&-A\lambda^2+
       \frac12 A\lambda^4-\Lambda \lambda^5e^{i\delta}&
         1-\frac12 A^2\lambda^4\cr }          
                 \right)+O(\lambda^6)\;,
\ena
where $\lambda=0.2205$, $A=0.6540$ and $\Lambda=1.100$ by using 
the mixing angles in Eq.(12). 

\subsection{ Explicit form of quark mass matrices at GUT scale}

Due to the hierarchy of magnitudes of quark masses, we 
parametrize $\tilde D_u$ and $\tilde D_d$ as follows, 
\bea
\tilde D_u=m_t\left (\matrix{\xi_{ut}\lambda^7& &\cr
                            & \xi_{ct}\lambda^4& \cr
                            &&1 \cr} \right )\;,
 \tilde D_d=m_b\left (\matrix{\xi_{db}\lambda^4& &\cr
                            & \xi_{sb}\lambda^2& \cr
                            &&1 \cr} \right )\;,
\ena 
where $\xi$'s are quantities of order unity. From Eq.(11), 
we have $|\xi_{ut}|=0.318$, $|\xi_{ct}|=0.990$,  
$|\xi_{db}|=0.563$ and $|\xi_{sb}|=0.545$.

By using them, one finds
\bea
\tilde M_d&=&m_b\left(\matrix{
     (\xi_{db}+\xi_{sb})\lambda^4&
      \xi_{sb}\lambda^3-(\xi_{db}+\frac12 \xi_{sb})\lambda^5
      &\Lambda\lambda^4e^{-i\delta}-\xi_{sb}A\lambda^5\cr
       \xi_{sb}\lambda^3-(\xi_{db}+\frac12 \xi_{sb})\lambda^5&
        \xi_{sb}\lambda^2+(-\xi_{sb}+A^2)\lambda^4&
         A\lambda^2-\xi_{sb}A\lambda^4\cr
         \Lambda\lambda^4e^{-i\delta}-\xi_{sb}A\lambda^5&
         A\lambda^2-\xi_{sb}A\lambda^4&1-A^2\lambda^4\cr
          }\right)\nonumber\\
         && +\left(\matrix{O(\lambda^6)&O(\lambda^6)&O(\lambda^7)\cr
         O(\lambda^6)&O(\lambda^6)&O(\lambda^6)\cr
         O(\lambda^7)&O(\lambda^6)&O(\lambda^6) }\right)
\ena

It is interesting to observe the difference between 
the mass hierarchy of u-type quarks and that of d-type quarks. 
While $|m_u/m_t|\sim O(\lambda^7)$ and $|m_c/m_t|\sim O(\lambda^4)$,  
 $|m_d/m_b|\sim O(\lambda^4)$ and $|m_s/m_b|\sim O(\lambda^2)$. 
 That is, the mass hierarchy of d-quarks is much less sevierer than 
 that of u-quarks. Next, we observe that $(\tilde M_d)_{22}\sim 
 (\tilde M_d)_{23}\sim O(\lambda^2)$. These are crucial in the 
 following discussions.

\subsection{The hierarchy in the neutrino mass matrix}
Firstly, we discuss what kind of neutrino mass matrix is required  
from the recent data. From the data[1],[2],[3], 
 the neutrino mass mixing matrix $O$ is almost 
fixed aside from CP violation phases as
\bea
O\sim\left( \matrix{1&\epsilon&\epsilon'\cr
                 -\epsilon&c & s\cr
                 \epsilon'&-s&c\cr}
\right )\;,
\ena 
where $s=\sin \theta_{\mu\tau}$ and $c=\cos \theta_{\mu\tau}$ with 
$\theta_{\mu\tau}$ is the mixing angle between $\nu_\mu$ and $\nu_\tau$ 
neutrinos, $\epsilon \sim \lambda^2$ and $\epsilon'/\epsilon\sim 
s/(1+c)$. From this mixing matrix, the expected neutrino mass 
matrix is given by
\bea
m_{\nu}=OD_{\nu}O^T\sim m_{\nu_\tau}\left( 
       \matrix{r_1+\epsilon'^2&s\epsilon'&c\epsilon'\cr
                 s\epsilon'&s^2 +r_2 c^2 &
                  sc \cr
                 c\epsilon'&sc 
                 &r_2 s^2+c^2\cr}
\right )\;,
\ena 
where $r_1=m_{\nu e}/m_{\nu\mu}$, 
$r_2=m_{\nu\mu}/m_{\nu\tau}\sim \pm 1/10$.  
If we take $\sin^2 2\theta_{\mu\tau} \ge 0.7$ for the experimental 
allowed region,  
we see that the submatrix relevant to $\nu_\mu$  and the $\nu_\tau$  
should have less hierarchical structure than  
quark mass matrices as  
\bea
     \left(\matrix{0.23+ 0.77r_2 &0.42
      \cr 0.42&0.23 r_2+0.77}\right) \sim
      \frac12 \left(\matrix{1+ r_2 &1
      \cr 1& r_2+1}\right)
      \sim \left(\matrix{0.77+0.23 r_2 
              &0.42\cr
              0.42&0.77 r_2+0.23}\right).\nonumber
              \\
\ena
The above matrices show variations of their components. They 
corresponds to the angle $\theta_{\mu\tau}=\theta_0$, $\pi/4$ and 
$\pi/2-\theta_0$ with $\sin^2 2\theta_0=0.7$.  From the above 
analysis, we observe the followings: (1) The 1-2 and 1-3 elements 
are of order $\lambda^2$, while the 1-1 element is of order $\lambda^4$. 
(2) In general, the non-hierarchical structure for the part 
related to $\nu_\mu$ and $\nu_\tau$. Even in the extreme case, 
the hierarchy is at most of order $\lambda$.

\subsection{The mechanism which leads the large mixing between $\nu_\mu$ and 
$\nu_\tau$}
 
In the basis where $\tilde M_u$ is diagonal,  $\tilde M_d$ takes 
a hierarchical form. Then, it is natural to suppose that $\tilde M_l$  
also take a hierarchical form. On the other hand, in order to 
get the non-hierarchical neutrino mass matrix 
for the part related to $\nu_\mu$ and $\nu_\tau$, at least 
one of $M_{\nu^D}$ and  $M_{\nu^R}$ should take the non-hierarchical 
form for the relevant part.  Since  $M_{\nu^D}$ and  $M_{\nu^R}$ 
are linear combinations of hierarchical mass matrices, 
$\tilde M_u$, $\tilde M_d$ and $\tilde M_l$,  
there needs some mechanism to get the non-hierarchical 
structure for $M_{\nu^D}$ and/or  $M_{\nu^R}$ for the relevant part. 
This is a necessary condition and does not imply the desired 
form of neutrio mass is obtained.  However, we seek this 
possibility. 

The hint lies in the fact that the 2-2 element of $\tilde M_d$ is the 
same size as the 2-3 element. 
We consider how to obtain the non-hierarchical 
form of $\tilde M_{\nu D}$.  By adjusting $r$, we make 
the 3-3 element of $\tilde M_{\nu D} = \tilde M_u - r \tilde M$ 
 as small as of order $\lambda^2 m_t$. Then, $m_c$ and $m_u$ do 
 not contribute to $\tilde M_{\nu D}$ because of the large 
 hierarchy of u-type quark masses. Thus,  only $m_t$ in 
 $\tilde M_{u}$ contributes to in $\tilde M_{\nu D}$. Thus 
 the non-hierarchical suturucture arises with the 
 above condition for the 3-3 element. 
 
We consider the above condition in detail, which is 
treated as two separate cases. 

\noindent
(a) $m_b m_\tau <0$ case

In this case, $\tilde M_{\nu R}$ has a hierarchical form as we see 
from Eqs.(8) and (9). We require that  
$\tilde M_{\nu D}$  has a non-hierarchical 
form for the relevant part. This is achieved by requiring that 
the 3-3 element of  $\tilde M_{\nu D}$ to be of order of the 2-3 element:
\bea 
 m_t-r(m_b-m_{\tau})\sim O(\lambda^{2} m_t)\;.
\ena
That is, the value of $r$ is fixed to be of order 
\bea
r\sim \frac{m_t}{m_b-m_\tau}\sim 50\;.
\ena
In particular, the range of $r$ which  gives a non-hierarchical 
neutrino mass structure will consist of two parts. One is the region 
 $r< 50$ and the other is the one $r>50$, because the exact 
 equality $m_t=r(m_b-m_\tau)$ gives the vanishing 3-3 element 
 of neutrino mass matrix so that we can not reproduce 
 the desired mass matrix given in Eq.(18). 

\noindent
(b) $m_b m_\tau >0$ case

In this case,  $\tilde M_{\nu R}$ has  a non-hierarchical form due to 
the cancellation between $m_b$ and $m_\tau$. There are two 
cases. 

\noindent 
(b-1) The case where $\tilde M_{\nu D}$ has  a non-hierarchical form:
\bea
 r\sim \frac{m_t}{m_b-m_\tau}\sim 750\;.
\ena
The range of $r$ which gives a non-hierarchical 
neutrino mass structure will consist of two parts, the one  
 $r< 750$ and the other $r>750$.

\noindent
(b-2) The case where $\tilde M_{\nu D}$ has  a hierarchical form:
\bea
  r<< \frac{m_t}{m_b-m_\tau}\;.
\ena
  
The above conditions are 
of course   only the necessary ones to achieve 
a non-hierarchical form of the neutrino mass matrix. The 
problem is whether there are regions of $r$ which 
reproduce the desired neutrino mixing angles and more 
importantly the hierarchical spectrum of neutrino masses. 
There is no  guarantee of the existence of such parameter 
region of $r$. We have to calculate the neutrino matrix 
for a given $r$ and examine whether the desired mixings and  
mass ratios are realized, especially by paying a special 
attention on the above mentioned parameter regions of $r$.

\section{The analysis }
Before the analysis, the following comments are in order:

\noindent
(i) The parameters

In addition to $r$ and $R$, we have sign freedoms of quark and lepton
masses. Since one of the phase 
can be fixed,  we choose $m_b>0$. If we scale $\tilde M_{\nu D}$ 
by $m_t$, the parameter $r$ enters as $r/m_t$. Thus, we can fix 
$r>0$, while we allow $m_t$ to take both positive and negative signs. 
Thus, we fix $m_b>0$ and $r>0$ and take all combinations of signs of 
other fermion masses. 

\noindent
(ii) The desired neutrino mixings and masses 

We consider  the following  
constraints on ranges of neutrino mixings and neutrino masses[1],[10]: 
\bea
3\times 10^{-3}< \sin^2 2\theta_{e\mu}< 2.0\times 10^{-2}\;,
& 0.7< \sin^2 2\theta_{\mu\tau}\;,\nonumber
\\
3\times 10^{-4}< \frac{\Delta m_{12}^2}{\Delta m_{23}^2}<
5\times 10^{-2}\;,&
\ena
where we used $3\times 10^{-6}{\rm eV}^2<
\Delta m_{12}^2<1\times 10^{-5}{\rm eV}^2$ and 
$2\times 10^{-4}{\rm eV}^2<
\Delta m_{23}^2<1\times 10^{-2}{\rm eV}^2$. 
Since we are dealing with the hierarchical mass spectrum of neutrino 
case, the overall normalization is fixed by the mass 
$m_{\nu\tau}$. The parameter $R$ is determined to fix $m_{\nu\tau}$ 
in the range of 
\bea
3\times 10^{-6}{\rm eV}^2< \Delta m_{12}^2  < 
1\times 10^{-5}{\rm eV}^2\;,\nonumber\\
2\times 10^{-4}{\rm eV}^2< \Delta m_{23}^2 < 
5\times 10^{-3}{\rm eV}^2\;.
\ena 
In the above, we defined $\Delta m^2_{jk}\equiv m_k^2-m_j^2$ and  
$m_1^2<<m_2^2<<m_3^2$.  

\noindent
(iii) The mixing angles 

Since we are looking for the solutions which reproduce 
the hierarchical neutrino mass spectrum  and 
the mixings in Eqs.(23) and (24), we may treat the three 
neutrino mixing as if it is due to the two neutrino mixing. 
In other words, we may define angles  by 
$\sin \theta_{e\mu}\simeq (U_\nu)_{12}$ and  
$\sin \theta_{\mu\tau}\simeq (U_\nu)_{23}$, 
where $U_\nu$ is the neutrino mixing matrix.  
With this approximation, we seek the range of parameter $r$ 
which reproduce the desired mixings and  ratios of masses to 
solve the atmospheric neutrino problem and the solar neutrino 
problem. Once we find the solution, then we examine whether 
 the mixings and masses that we obtained do not violate the 
 CHOOZ bound for the $\bar \nu_e-\bar\nu_X$ oscillation[2]. Since 
 $\Delta m_{12}^2<<\Delta m_{23}^2\simeq \Delta m_{13}^2\simeq m_3^2$, 
 we can 
treat the mixing angle as $\sin \theta_{eX}=(U_\nu)_{13}$ and the mass 
 squared difference as $\Delta m^2\simeq m_3^2$ in our three neutrino 
 mixing scenario. 
 
\subsection{The simplification of the problem}

In this paper, we use the following simplification: 

\noindent
(i) The CP violation is neglected. Explicitly, we perform 
the numerical analysis by setting the Kobayashi-Maskawa CP 
violation angle $\delta$ zero and taking quark and charged 
lepton masses are real. However, we set signs of 
fermion masses are free so that we have to consider all 
combinations of signs of quark and charged lepton masses. 

\noindent
(ii) $\tilde M_l$ is assumed to be diagonal in the basis we 
adopted in this paper.  

With these simplifications, neutrino mass matrix 
$m_\nu$ is determined by quark and charged lepton masses,  
CKM mixing angles, $r$ and $R$. The parameter $R$ determines 
the overall scale of $\tilde M_{\nu R}$ so that it plays  a role of 
adjusting the overall scale of neutrino masses in the see-saw 
mechanism. The neutrino mixing angles and the ratios of 
neutrino masses 
are solely determined  by  only one parameter $r$. 
Thus, the  present model is quite tightly constrained  one. 
 
Once the ranges of parameter $r$ which reproduce the desired 
neutrino mixing angles and ratios of masses are found, 
 the introduction of CP violation  will 
relax the region of $r$ and $R$.  The same is expected if we 
relax (ii). 
Therefore, in this paper we concentrate 
on finding the region of $r$ and $R$ in the simplified and 
tight situation. 

\subsection{The result}

The procedure of our analysis is as follows: For a given $r>0$ 
and a combination of signs of fermion masses, 
$\tilde M_{\nu D}$ and $\tilde M_{\nu R}/R$ are calculated. 
We  compute the neutrino mass matrix by the see-saw 
mechanism.  Then, we  compute neutrino mixings and neutrino mass ratios 
to see whether the result reproduce the desired ones. .

We are fortunate that we found  many regions of $r$ 
which reproduce the desired mixings and masses. 
The result is summarized in Table 1 and 2. 

In Table 1a, we show the result for $m_bm_\tau<0$. Solutions 
exist for various  combinations of signs of quark and charged lepton as 
we see from the  table.  Each sign combination is expressed by the following 
abbreviated notation, Ei-Dj-UkX. The indices i, j and k runs  1, 2, 3 and 4 
which express sings of masses as $(-,-)$, $(-,+)$, $(+,-)$ and $(+,+)$ 
for a pair $(e,\mu)$ 
(denoted by E), $(d,s)$ (denoted by D) and $(u,c)$ (denoted by U), 
respectively, and X takes P and M that means the sign of $m_t$ is 
positive and negative, respectively. The region of $r$ 
which reproduces the data in eqs.(23) is given in the column 
$r$. the column $\sin^2 2\theta_{\mu\tau}$ show the 
largest value of it, which is achieved by a specific value of $r$ 
shown in the column $R(10^{13})(r)$. The value of $R$ fixes the 
overall normalization of neutrino mass, especially $m_{\nu\tau}$. 
We found solutions only for  $m_\tau<0$  in  the range  of 
$r=54 \sim 64$ which coincide with our expectation given in Eq.(20). 
For all cases $\tilde M_{\nu D}$ has a
non-hierarchical form, while $\tilde M_{\nu R}\equiv R\tilde M$ takes 
a  hierarchical form. 
 The mixings and masses with specific values of $r$ and $R$ 
 for all cases are given in Table 1b. 
 
In Table 2a, we show the result for $m_bm_\tau>0$ case. The notation 
to discriminate models which differ by sign combinations of fermion 
masses is the same as in Table 1a. 
We see that for all solutions $m_t<0$ and 
 $r\sim (500\sim 960)$, except one case (E1-D2-U2P). These cases 
coincide what we expected from Eq.(21) and  both $\tilde M_{\nu D}$ and  
$\tilde M_{\nu R}\equiv R\tilde M$ take non(less)-hierarchical forms.

The case  E1-D2-U2P is realized when  $r \sim 35$, which is also 
what we expected in Eq.(22), although the value of $r$ is unknown.  
 In this case, $\tilde M_{\nu D}$ has a hierarchical 
form, while  $\tilde M_{\nu R}\equiv R\tilde M$ takes a 
non(less)-hierarchical form. 
In Table 2b, we show the mixings and masses for all cases.

For completeness, the neutrino 
mass matrix and the neutrino mixing matrix for all cases 
are given in Appendix B, except for some specific cases which 
we show below to see the details of the models.  

\noindent
(1) $m_bm_\tau<0$ case: 

\noindent
(1-1) E1-D3-U2P  with $r=58.2$ 

The Dirac mass term and the Majorana mass term are given by 
\bea
\tilde M_{\nu D}=\left(\matrix{-0.0190&-0.3530&-0.1402\cr
                      -0.3530&-2.2871&-1.8954\cr
                      -0.1402&-1.8954&2.6852\cr}  \right ),\;
\tilde M_{\nu R}={R}\left(\matrix{0.0003&0.0061&0.0024\cr
                      0.0061&0.0445&0.0326\cr
                      0.0024&0.0326&2.1704\cr}  \right ).
                      \nonumber\\
                      \ena
As we expected, $M_{\nu D}$ has a non-hierarchical form the part 
relevant to the second and the third generations, while $M_{\nu R}$ is 
hierarchical, because there is no cancellation of the 3-3 element of 
$M_{\nu R}$. 
The neutrino mass matrix and the neutrino mixing matrix is 
given by 
\bea
m_\nu=-\frac 1{R}\left(\matrix{1.165&20.46&8.402\cr
                      20.46&114.3&106.7\cr
                      8.402&106.7&59.82\cr}  \right ),\;
 U_\nu=\left(\matrix{0.964&0.067&-0.256\cr
                      -0.242&0.613&-0.752\cr
                      0.107&0.787&0.608\cr}  \right ).
\ena
Then, we have $\sin^2 2\theta_{e\mu}=0.018$ and 
$\sin^2 2\theta_{\mu\tau}=0.98$. With $R=1.09\times 10^{13}$, 
we found 
\bea
m_1=-3.22\times 10^{-5}{\rm eV}\;, \;m_2=2.25\times 10^{-3}{\rm eV}\;,
 \;m_3=-1.83\times 10^{-2}{\rm eV}\;, 
\ena
which are multiplied by $1.09\times 10^{13}/R$ if the $R$ dependence is  
kept.  Thus
\bea
\Delta m_{12}^2=5.08\times 10^{-6}{\rm eV^2}\;,
\; \Delta m_{23}^2=3.30\times 10^{-4}{\rm eV^2}\;. 
\ena

As we have mentions before, the mixing angles are computed by 
assigning $\sin \theta_{e\mu}=(U_\nu)_{12}$ and 
$\sin \theta_{\mu\tau}=(U_\nu)_{23}$. This approximation is  
 reasonable for the mixings and the hierarchical mass spectrum 
 given in Eqs.(26) and (27). 
 
The sensitivity of mixing angles $\sin \theta_{\mu\tau}$ and 
$\sin \theta_{e\mu}$ and also the ratio $\Delta m_{12}^2/\Delta m_{23}^2$ 
with respect to the parameter $r$ is shown in Figs.1a, 1b and 1c, 
respectively. The angle $\sin \theta_{\mu\tau}$ takes values larger 
than 0.7 for a wider range of $r$. The same holds for the ratio 
$\Delta m_{12}^2/\Delta m_{23}^2$. The allowed region of $r$ is essentially 
fixed by $\sin \theta_{e\mu}$ and the region is between 58 to 62. 
This situation holds for all cases E1-D3-UiP $(i=1\sim 4)$. We will 
show the comparison with CHOOZ data, the disappearance test of $\bar{\nu}_e$, 
 at the end of this section. 

\noindent
(1-2) E4-D2-U3P  with $r=58.1$ 

The Dirac mass term and the Majorana mass term are similar to 
the case (1-1), so that we give only  the neutrino mass matrix 
and the neutrino mixing matrix 
\bea
m_\nu=\frac 1{R}\left(\matrix{1.115&19.83&-9.592\cr
                      19.83&107.0&-98.81\cr
                      -9.592&-98.81&60.02\cr}  \right ),
 U_\nu=\left(\matrix{0.960&0.0708&0.270\cr
                      -0.254&0.620&0.742\cr
                      -0.115&-0.781&0.614\cr}  \right ).
\ena     
Then, we have $\sin^2 2\theta_{e\mu}=0.0200$ and 
$\sin^2 2\theta_{\mu\tau}=0.990$. With $R=0.865\times 10^{13}$, 
we found 
\bea
m_1=-1.33\times 10^{-5}{\rm eV}\;, \;m_2=-2.23\times 10^{-3}{\rm eV}\;,
 \;m_3=2.17\times 10^{-2}{\rm eV}\;, 
\ena
which are multiplied by $0.865\times 10^{13}/R$ if the $R$ dependence is  
kept.  Thus
\bea
\Delta m_{12}^2=4.97\times 10^{-6}{\rm eV^2}\;,
\; \Delta m_{23}^2=4.05\times 
10^{-4}{\rm eV^2}\;. 
\ena
 
As we can see from Figs.2a, 2b and 2c, 
the angle $\sin \theta_{\mu\tau}$ takes values larger 
than 0.7 for a wider range of $r$. The same holds for the ratio 
$\Delta m_{12}^2/\Delta m_{23}^2$. The allowed region of $r$ is essentially 
fixed by $\sin \theta_{e\mu}$. There are three allowed regions. 
This situation holds for all cases E2-D2-UiP, E2-D4-UiP, E4-D2-UiP,
and E2-D4-UiP $(i=1\sim 4)$.  

\noindent
(2) $m_bm_\tau>0$ case:

\noindent
(2-1) E1-D3-U4M  with $r=680$ 

The Dirac mass term and the Majorana mass term are given by 
\bea
\tilde M_{\nu D}=\left(\matrix{-0.2089&-4.124&-1.638\cr
                      -4.124&-29.95&-22.15\cr
                      -1.638&-22.15&-11.74\cr}  \right ),\;
\tilde M_{\nu R}={R}\left(\matrix{0.0003&0.0061&0.0024\cr
                      0.0061&0.0445&0.0326\cr
                      0.0024&0.0326&-0.172\cr}  \right ).
                      \nonumber\\
\ena
As we expected, both $M_{\nu D}$ and $M_{\nu R}$ have  
non-hierarchical forms for the part 
relevant to the second and the third generations.  
The neutrino mass matrix and the neutrino mixing matrix is 
given by 
\bea
m_\nu=\frac1{R}\left(\matrix{-141.3&-2805&-1111\cr
                      -2805&-20159&-15018\cr
                      -1111&-15018&-7336\cr}  \right ),
 U_\nu=\left(\matrix{0.968&0.0485&-0.246\cr
                      -0.231&0.553&-0.800\cr
                      0.0973&0.831&0.547\cr}  \right ).  
\ena

Then, we have $\sin^2\theta_{e\mu}=0.009$ and 
$\sin^2\theta_{\mu\tau}=0.92$. With $R=1.22\times 10^{15}$, 
we found 
\bea
m_1=2.21\times 10^{-7}{\rm eV}\;, \; m_2=2.24\times 10^{-3}{\rm eV}\;,
 \; m_3=-2.49\times 10^{-2}{\rm eV}\;. 
\ena 
Thus
\bea
\Delta m_{12}^2=5.00\times 10^{-6}{\rm eV^2}\;, 
\;\Delta m_{23}^2=16.15\times 
10^{-4}{\rm eV^2}\;. 
\ena

As we can see from Figs.3a, 3b and 3c, 
the angle $\sin \theta_{\mu\tau}$ takes values larger 
than 0.7 for a wider range of $r$. The same holds for the ratio 
$\Delta m_{12}^2/\Delta m_{23}^2$. The allowed region of $r$ is essentially 
fixed by $\sin \theta_{e\mu}$. There are three allowed regions 
similarly to the case (1-1). 
This situation holds for all cases E1-D3-UiM $(i=1\sim 4)$. 

\noindent
(2-2) E3-D3-U3M  with $r=790$ 

The Dirac mass term and the Majorana mass term are  similar to 
the case (2-1) so that we only show 
the neutrino mass matrix and the neutrino mixing matrix
\bea
m_\nu=\frac1{R}\left(\matrix{214.7&-3785&-1502\cr
                      -3785&-28242&-20425\cr
                      -1502&-20425&-9864\cr}  \right ),
 U_\nu=\left(\matrix{0.962&0.0630&-0.267\cr
                      -0.257&0.544&-0.799\cr
                      0.0946&0.837&0.539\cr}  \right ).  
\ena

As we can see from Figs.4a, 4b and 4c, 
the angle $\sin \theta_{\mu\tau}$ takes values larger 
than 0.7 for a wider range of $r$. The same holds for the ratio 
$\Delta m_{12}^2/\Delta m_{23}^2$. The allowed region of $r$ is essentially 
fixed by $\sin \theta_{e\mu}$. There are two allowed regions. 
This situation holds for all cases E1-D1-UiM, E3-D1-UiM, E3-D3-UiM 
$(i=1\sim 4)$. 

\noindent 
(2-3) E1-D2-U2P  with $r=36.5$ 

This is a very special case corresponding the case (b-2). 
The Dirac mass term and the Majorana mass term are given by 
\bea
\tilde M_{\nu D}&=&\left(\matrix{-0.01399&0.21534&-0.10180\cr
                      0.21534&-3.1557&-1.1315\cr
                      -0.10180&-1.1315&135.29\cr}  \right ),\nonumber\\
\tilde M_{\nu R}&=&{R}\left(\matrix{0.0003549&-0.0058998&0.002789\cr
                      -0.0058998&0.094732&0.031000\cr
                      0.002789&0.031000&-0.17239\cr}  \right ).
\ena
As we expected, $M_{\nu D}$ has a hierarchical form the part 
relevant to the second and the third generations, while $M_{\nu R}$ is 
less hierarchical, because of the  cancellation of the 33 element in 
$M_{\nu R}$. 
The neutrino mass matrix and the neutrino mixing matrix is 
given by 
\bea
m_\nu=\frac 1{R}\left(\matrix{-0.559&8.01&28.0\cr
                      8.01&-107&-606\cr
                      28.0&-606&-800\cr}  \right ),\;
 U_\nu=\left(\matrix{0.999&0.0374&-0.00658\cr
                     -0.0290&0.864&-0.502\cr
                      -0.007&0.502&0.865\cr}  \right ).
\ena
Then, we have $\sin^2\theta_{e\mu}=0.006$ and 
$\sin^2\theta_{\mu\tau}=0.75$. With $R=7.85\times 10^{13}$, 
we found 
\bea
m_1=-9.47\times 10^{-7}{\rm eV}\;, \; m_2=3.12\times 10^{-3}{\rm eV}\;, 
\; m_3=-1.47\times 10^{-2}{\rm eV}\;. 
\ena
  Thus
\bea
\Delta m_{12}^2=9.71\times 10^{-6}{\rm eV^2}\;,
\; \Delta m_{23}^2=2.06\times 
10^{-4}{\rm eV^2}\;. 
\ena

As we can see from Figs.5a, 5b and 5c, $\sin \theta_{e\mu}$ is 
insensitive to $r$ and the allowed region is determined by 
 $\sin \theta_{\mu\tau}$ and $\Delta m_{12}^2/\Delta m_{23}^2$. 
 The allowed region of $r$ is a tiny region,  so that the 
 solution is quite sensitive to the value of $r$.  This case 
 will be the most unlikely one.

Finally, we comment on the CHOOZ data[2]. Since 
$\Delta m_{12}^2<<\Delta m_{23}^2$, we can consider 
the mixing angle for the CHOOZ disappearance test of $\nu_e$ 
may be defined by $\sin \theta_{eX}=(U_\nu)_{13}$ which is 
$\sin \theta_{e\tau}$ in our cases and the 
mass squared difference is $\Delta m^2 \simeq m_{\nu\tau}^2 \equiv 
m_3^2$. Then, we compared these values of parameters with the 
CHOOZ two neutrino analysis. We found the followings: All 
cases except E1-D2-U2P predict $\sin 2\theta_{e\tau}\sim (0.2\sim 0.3)$. 
For $m_b m_\tau<0$ with $m_\mu m_c<0$ cases and also $m_bm_\tau>0$ cases, 
we predict $\Delta m^2<10^{-3}{\rm eV}^2$, so that they are safe. 
For $m_b m_\tau<0$ with $m_\mu m_c>0$ cases, we predict  
$\Delta m^2\sim 2\times 10^{-3}{\rm eV}^2$ and their values are on the 
boundary of the excluded region. This can be remedied by changing $r$ 
slightly from the value we took to achieve the largest value of 
$\sin 2\theta_{\mu\tau}$ or by taking larger value of $R$ to reduce the 
overall neutrino mass scale. For the E1-D2-U2P case, we 
predict $\sin 2\theta_{e\tau}\sim 1\times 10^{-4}$, so that this 
 satisfies the bound. In summary, many of our cases predict 
in general rather large values of $(U_\nu)_{13}\sim 0.25$. However, 
all cases satisfy the CHOOZ bound. In several cases corresponding 
to $m_bm_\tau<0$ with $m_\mu m_c>0$, the $\nu_e-\nu_\tau$ oscillation 
is large enough to be observed in the near future experiments. 
  
We explicitly showed  some of our solutions. Other solutions 
show the similar matrices to one of the above cases. We listed 
the neutrino mass matrix $m_\nu$ and the neutrino mixing matrix 
$U_\nu$ in Appendix B for some of other cases.

\section{The summary}
We showed the mechanism to induce the non-hierarchical neutrino 
mass matrix by using the hierarchical forms of mass matrices 
of $\tilde M_u$, $\tilde M_d$ and $\tilde M_l$, although it is  
a quite non trivial problem. Our model contains only one adjustable 
parameter $r$ which determines neutrino mixings and the ratios of 
neutrino masses. The other parameter $R$ is used to determine 
the absolute magnitude of neutrino masses, so that it determines 
 $m_{\nu\tau}$ in our hierarchical mass spectrum of neutrinos. 

Following our mechanism, we found many solutions, which are 
classified into three cases. Our mechanism guarantees  the 
non-hierarchical structure  only for the part relevant 
to the second and the third generations and thus we inevitably predict 
the small mixing between $\nu_e$ and $\nu_\mu$, while the large 
mixing between  $\nu_\mu$ and $\nu_\tau$.  We examined the $r$ dependence 
of mixing angles and the ratios of neutrino masses and show 
the sensitivity of these quantities to $r$. Depending on the 
choice of relative signs of fermion masses, the pattern of 
sensitivity changed. We showed the solutions are  not very 
sensitive to $r$ except the case E1-D2-U2P case.

There arises a question whether the existence of the solutions 
depends crucially on the values of quark masses and quark mixings, 
i.e., the value of $\tan \beta$. We analyzed by taking another 
set of values and found that the solutions given in this paper 
exist by the small change of the values of $r$ and $R$. Thus 
the existence of our solutions does not depend on them. 
 
Finally, we comment on the CP violation. In the present analysis, 
we ignored the CP violation effects. When the CP violation turns 
on, many phases enter in our model. One is  from the 
Kobayashi-Maskawa phase and others are from the phases of quark and 
charged lepton masses. These phases will relax the tight situation 
which we considered. Thus, we expect that with the CP violation 
models cover the broader range of the mixing angles and the 
neutrino masses. This problem is under considerations. 

Also, there is no good reason to assume that $\tilde M_l$ is 
diagonal, although it would be a hierarchical form. 
If we relax this assumption, the models would cover wider range of 
mixings and masses than what we obtained in this paper.  

\vskip 1cm
{\Huge Acknowledgment}

 This work is supported in part by 
the Japanese Grant-in-Aid for Scientific Research of
Ministry of Education, Science, Sports and Culture, 
No.08640374 and No.10140216.
  
\newpage

%Appendix format setting
\setcounter{section}{0}
\renewcommand{\thesection}{\Alph{section}}
\renewcommand{\theequation}{\thesection .\arabic{equation}}
\newcommand{\apsc}[1]{\stepcounter{section}\noindent
\setcounter{equation}{0}{\Large{\bf{Appendix\,\thesection:\,{#1}}}}}

\apsc{The proof of the incompatibility of the minimal model and 
quark and charged lepton mass spectrum }

For one 10 dimensional Higgs field $n=1$ case. we have additional 
relation, 
\bea
\tilde M_l = c_u \tilde D_u + c_d \tilde M_d\;,
\ena
where 
\bea
c_u=\left( 1-\frac 4{1-\frac{\kappa_d}{\kappa_u}\frac{v_u}{v_d}}\right)
\;,
c_d=4\left(\frac{\frac{v_u}{v_d}}
{1-\frac{\kappa_d}{\kappa_u}\frac{v_u}{v_d}}\right)\;.
\ena

In this section, we first discuss that the model contains one 10 and 
one 126 representation can not reproduce the charged lepton masses 
and thus the model is rejected independently of neutrino masses and 
mixings. Then, we go to the model which contains n $(\ge 2)$ 10 and 
one 126 models. We consider the ordinary see-saw mechanism to 
derive the left-handed neutrino mass matrix. In the case of type II 
see-saw model proposed by Mohapatra and Senjanovic[10], there appears the 
left-handed neutrino mass matrix. In our case, we do not consider 
this by taking $v_L$ is small. 

This model was considered by Mohapatra and his coworkers[4],[5],[8] 
as a model to give  
an unified model to explain quark and lepton masses and mixings. 
Recently, Brahmachari and Mohapatra[8] showed that this model 
can not explain 
the desired neutrino mixing pattern, the small mixing between 
$\nu_e$ and $\nu_\mu$, and the large mixing between 
$\nu_\mu$  and $\nu_\tau$. 

Here, we show that this model is not able to explain the charged 
lepton masses, so that the model is rejected independently 
of the arguments on neutrino masses and mixings. Firstly, we 
see that the charged lepton mass matrix is written in a linear 
combination of  quark mass matrices as in Eq.(A.1).

Because $|m_\tau|\simeq |m_b|\sim \lambda^3 |m_t|$ with 
the the Cabibbo angle $\lambda=0.2205$, $c_u \sim \lambda^3$ 
and $c_d\sim 1$. Knowing that $\tilde D_u$ is a diagonal matrix and 
$\tilde M_d$ is almost diagonal, and also the mass hierarchy of u-quark 
sector is much severer that that of d-quark, we observe that 
the contribution to the first and the second generation part of 
$\tilde M_l$ from the u-quark part $D_u$ is negligible so that it is 
proportional to that of $\tilde M_d$. Thus, the mass matrix 
that predicts $m_e/m_\mu\simeq m_d/m_s$ 
does not reproduce the observed hierarchical structure of d-quark 
and the charged lepton masses such as predicted by 
the Georgi-Jarskog mass 
relations $m_b=m_\tau$, $m_s=m_\mu/3$ and $m_d=3m_e$ at GUT 
scale. 

\vskip 1cm
\apsc{The expressions of neutrino mass matrix and the neutrino 
mixing matrix}

Here, we show the explicit forms of the neutrino mass matrix $m_\nu$ 
and the neutrino mixing matrix $U_\nu$ for various cases. 
From these matrices, the mixing angles and the ratios of neutrino masses 
are obtained. The result is summarized in Table 1b and 2b.

\subsection{ $m_bm_\tau<0$ case}

\noindent
The case E2-D2-U1P  with $r=56.1$ predicts 
\bea
m_\nu=\frac1{R}\left(\matrix{-1.234&18.60&-8.698\cr
                      18.60&100.4&-87.78\cr
                      -8.698&-87.78&48.17\cr}  \right ),
 U_\nu=\left(\matrix{0.957&0.0686&0.283\cr
                      -0.266&0.599&0.755\cr
                      -0.118&-0.796&0.592\cr}  \right ).
\ena
\vskip 5mm 

\noindent
The case E2-D4-U1P  with $r=62.9$ predicts
\bea
m_\nu=\frac1{R}\left(\matrix{-11.55&21.09&-10.92\cr
                      21.09&131.2&-110.3\cr
                      -10.92&-110.3&52.03\cr}  \right ),
 U_\nu=\left(\matrix{0.963&0.0601&0.263\cr
                      -0.248&0.578&0.778\cr
                      -0.105&-0.814&0.571\cr}  \right ).
\ena                      
\vskip 5mm 
                      
\noindent
The case E4-D4-U1P  with $r=55.3$ predicts
\bea
m_\nu=\frac 1{R}\left(\matrix{-6.952&16.31&-8.429\cr
                      16.31&97.70&-82.65\cr
                      -8.429&-82.65&40.67\cr}  \right ),
 U_\nu=\left(\matrix{0.962&0.0594&0.266\cr
                      -0.250&0.583&0.773\cr
                      -0.109&-0.810&0.576\cr}  \right ).
\ena                      

\subsection{The $m_bm_\tau>0$ case}

\noindent
(2-1) E1-D1-U2M  with $r=830$ 
  
\bea
m_\nu=\frac1{R}\left(\matrix{1529&-3784&-1677\cr
                      -3784&-30058&-22307\cr
                      -1677&-22307&-10215\cr}  \right ),\;
 U_\nu=\left(\matrix{0.965&0.0634&-0.254\cr
                      -0.247&0.545&-0.801\cr
                      0.0877&0.836&0.542\cr}  \right ).
\ena
\vskip 5mm

\noindent
The case E3-D1-U2M  with $r=840$ predicts 
\bea
m_\nu=\frac1{R}\left(\matrix{2025&-3876&-1717\cr
                      -3876&-30793&-22848\cr
                      -1717&-22848&-9985\cr}  \right ),
 U_\nu=\left(\matrix{0.964&0.0654&-0.258\cr
                      -0.252&0.541&-0.802\cr
                      0.0872&0.838&0.538\cr}  \right ).  
\ena

%%%%%%%%%%%%%%%%%%%%%%%%%%%%%%%%%%%%%%%%%%%%%%%%%%%%%%%%%%%%%%%%%
\newpage
\renewcommand{\oddsidemargin}{-3cm}
\renewcommand{\evensidemargin}{-3cm}
\renewcommand{\arraystretch}{0.8}
\normalsize
\noindent
Table 1a: The ranges of parameters of $r$ and $R$ to predict 
the desired values of neutrino mixing angles and masses 
for $m_bm_\tau<0$. 
\\
\vskip 5mm
\scriptsize 
E1-D3-UiP (i=1,2,3,4)

\begin{tabular}{c|c||c||c|c|c}
\hline \hline
$ e , \mu$ & $ d , s$ & 
$ u , c , t $ & $r$ & $R$ $(10^{13})$ $(r)$&
$\sin^2{2\theta_{\mu\tau}}$\\ \hline \hline
$(-,-)$&$(+,-)$
&$(-,-),+$&58.9-60.9&0.62-0.73 (60.9)&0.89\\
&&$(-,+),+$&58.2-61.7&0.78-1.4 (58.2)&0.98\\
&&$(+,-),+$&59.3-60.5&0.61-0.78 (60.5)&0.89\\
&&$(+,+),+$&58.5-61.4&0.83-1.4 (58.5)&0.98\\
\hline \hline
\end{tabular}

\vspace{10mm}
E2-D2-UiP (i=1,2,3,4)

\begin{tabular}{c|c||c||c|c|c}
\hline \hline
$ e , \mu$ & $ d , s$ & 
$ u , c , t $ & $r$ & $R$ $(10^{13})$ $(r)$&
$\sin^2{2\theta_{\mu\tau}}$\\ 
\hline \hline
$(-,+)$&$(-,+)$&$(-,-),+$&
55.0, 55.8-56.1, 61.9-62.3, 63.3-63.9&0.58-1.05 (56.1)&0.98\\
&&$(-,+),+$&
55.0-55.6, 56.4-56.6, 61.4-61.7, 62.6-63.8&0.62-0.73 (61.4)&0.89\\
&&$(+,-),+$&
55.0-55.1, 55.8-56.1, 61.8-62.2, 63.2-63.9&0.58-1.05 (56.1)&0.98\\
&&$(+,+),+$&
55.0-55.7, 56.5-56.7, 61.3-61.6, 62.5-63.7&0.62-0.69 (61.3)&0.90\\
\hline \hline
\end{tabular}

\vspace{10mm}
E2-D4-UiP (i=1,2,3,4)

\begin{tabular}{c|c||c||c|c|c}
\hline \hline
$ e , \mu$ & $ d , s$ & 
$ u , c , t $ & $r$ & $R$ $(10^{13})$ $(r)$&
$\sin^2{2\theta_{\mu\tau}}$\\ 
\hline \hline
$(-,+)$&$(+,+)$&$(-,-),+$&
55.0-55.1, 62.9-63.1, 63.8-64.4&0.73-1.32 (62.9)&0.96\\
&&$(-,+),+$&
55.0-55.2, 55.8, 62.3-62.4, 63.0-64.1&0.64-0.81 (62.3)&0.86\\
&&$(+,-),+$&
55.1-55.2, 62.9-63.1, 63.8-64.4&0.54-0.97 (55.2)&0.96\\
&&$(+,+),+$&
55.0-55.3, 55.8-55.9, 62.2-62.3, 63.0-64.1&0.64-0.75 (62.2)&0.89\\
\hline \hline
\end{tabular}

\vspace{10mm}
E4-D2-UiP (i=1,2,3,4)

\begin{tabular}{c|c||c||c|c|c}
\hline \hline
$ e , \mu$ & $ d , s$ & 
$ u , c , t $ & $r$ & $R$ $(10^{13})$ $(r)$&
$\sin^2{2\theta_{\mu\tau}}$\\ 
\hline \hline
$(+,+)$&$(-,+)$&$(-,-),+$&
55.3-55.6, 56.8-57.7, 60.2-61.2, 62.6-63.2&0.65-1.17 (57.7)&0.99\\
&&$(-,+),+$&
55.4-56.2, 57.2-57.6, 60.4-60.8, 61.9-63.1&0.38-0.68 (60.8)&0.85\\
&&$(+,-),+$&
55.3-55.8, 57.0-58.1, 59.8-61.0, 62.4-63.2&0.62-1.11 (58.1)&0.99\\
&&$(+,+),+$&
55.4-56.4, 57.4-57.8, 60.2-60.7, 61.8-63.1&0.36-0.64 (60.7)&0.85\\
\hline \hline
\end{tabular}

\vspace{10mm}
E4-D4-UiP (i=1,2,3,4)

\begin{tabular}{c|c||c||c|c|c}
\hline \hline
$ e , \mu$ & $ d , s$ & 
$ u , c , t $ & $r$ & $R$ $(10^{13})$ $(r)$&
$\sin^2{2\theta_{\mu\tau}}$\\ 
\hline \hline
$(+,+)$&$(+,+)$&$(-,-),+$&
55.2-55.3, 62.7-62.9, 63.6-64.3&0.54-0.98 (55.3)&0.96\\
&&$(-,+),+$&
55.0-55.4, 55.9-56.0, 62.0-62.2, 62.8-64.0&0.64-0.72 (62.0)&0.89\\
&&$(+,-),+$&
55.2-55.3, 62.6-62.9, 63.6-64.3&0.67-1.21 (62.6)&0.97\\
&&$(+,+),+$&
55.0-55.4, 56.0-56.1, 62.0-62.1, 62.8-64.0&0.53-0.58 (56.1)&0.89\\
\hline \hline
\end{tabular}

%%%%%%%%%%%%%%%%%%%%%%%%%%%%%%%%%%%%%%%%%%%%%%%%%%%%%%%%%%%%%%%%%%%
\newpage
\normalsize
\noindent
Table 1b: Predictions of neutrino mixing angles and masses 
for $m_bm_\tau<0$. All cases correspond to Table 1a. 
\\
\vskip 3mm
\scriptsize
\begin{tabular}{c|c|c|c|c|c|c|c}
\hline \hline
Model &
\lw{$\sin^2{2\theta_{e\mu}}$}&      %% atmospheric
\lw{$\sin^2{2\theta_{\mu\tau}}$} &  %% solar (MSW small)
\lw{$m_1$($10^{-3}$eV)}&            %% m(nu_e)
\lw{$m_2$($10^{-3}$eV)}&            %% m(nu_mu)
\lw{$m_3$($10^{-2}$eV)}&            %% m(nu_tau) 
\lw{$\Delta m_{12}^2$(eV$^2$)}&
\lw{$\Delta m_{23}^2$(eV$^2$)} \\
$(r , R/10^{13})$ & & & & & & & \\ \hline \hline
E1-D3-U1P &
\lw{0.019} & \lw{0.89} & \lw{$-0.0250$} &
\lw{$1.89$} & \lw{$-4.08$} &
\lw{$3.56\times10^{-6}$} & \lw{$1.66\times10^{-3}$}\\ 
(60.9, 0.675) & & & & & & & \\ \hline
E1-D3-U2P &
\lw{0.018} & \lw{0.98} & \lw{$-0.0322$} &
\lw{$2.25$} & \lw{$-1.83$} &
\lw{$5.08\times10^{-6}$} & \lw{$3.30\times10^{-4}$}\\ 
(58.2, 1.09) & & & & & & & \\ \hline
E1-D3-U3P &
\lw{0.020} & \lw{0.89} & \lw{$-0.00946$} &
\lw{$1.97$} & \lw{$-3.91$} &
\lw{$3.87\times10^{-6}$} & \lw{$1.52\times10^{-3}$}\\ 
(60.5, 0.695) & & & & & & & \\ \hline
E1-D3-U4P &
\lw{0.019} & \lw{0.98} & \lw{$-0.0231$} &
\lw{$2.35$} & \lw{$-1.80$} &
\lw{$5.51\times10^{-6}$} & \lw{$3.19\times10^{-4}$}\\ 
(58.5, 1.115) & & & & & & & \\ \hline 
\hline
E2-D2-U1P &
\lw{0.019} & \lw{0.98} & \lw{$-0.304$} &
\lw{$-2.26$} & \lw{$2.06$} &
\lw{$5.03\times10^{-6}$} & \lw{$4.21\times10^{-4}$}\\ 
(56.1, 0.815) & & & & & & & \\ \hline
E2-D2-U2P &
\lw{0.014} & \lw{0.89} & \lw{$-0.436$} &
\lw{$-1.93$} & \lw{$4.09$} &
\lw{$3.54\times10^{-6}$} & \lw{$1.67\times10^{-3}$}\\ 
(61.4, 0.675) & & & & & & & \\ \hline
E2-D2-U3P &
\lw{0.015} & \lw{0.98} & \lw{$-0.289$} &
\lw{$-2.25$} & \lw{$2.06$} &
\lw{$4.98\times10^{-6}$} & \lw{$4.21\times10^{-4}$}\\ 
(56.1, 0.815) & & & & & & & \\ \hline
E2-D2-U4P &
\lw{0.017} & \lw{0.90} & \lw{$-0.411$} &
\lw{$-1.87$} & \lw{$4.21$} &
\lw{$3.33\times10^{-6}$} & \lw{$1.77\times10^{-3}$}\\ 
(61.3, 0.655) & & & & & & & \\ \hline 
\hline
E2-D4-U1P &
\lw{0.014} & \lw{0.96} & \lw{$-1.29$} &
\lw{$-2.58$} & \lw{$2.06$} &
\lw{$4.98\times10^{-6}$} & \lw{$4.18\times10^{-4}$}\\ 
(62.9, 1.025) & & & & & & & \\ \hline
E2-D4-U2P &
\lw{0.010} & \lw{0.86} & \lw{$-1.77$} &
\lw{$-2.63$} & \lw{$3.92$} &
\lw{$3.80\times10^{-6}$} & \lw{$1.53\times10^{-3}$}\\ 
(62.3, 0.725) & & & & & & & \\ \hline
E2-D4-U3P &
\lw{0.018} & \lw{0.96} & \lw{$-1.32$} &
\lw{$-2.59$} & \lw{$2.08$} &
\lw{$4.97\times10^{-6}$} & \lw{$4.28\times10^{-4}$}\\ 
(55.2, 0.755) & & & & & & & \\ \hline
E2-D4-U4P &
\lw{0.019} & \lw{0.89} & \lw{$-1.80$} &
\lw{$-2.60$} & \lw{$4.09$} &
\lw{$3.53\times10^{-6}$} & \lw{$1.66\times10^{-3}$}\\ 
(62.2, 0.695) & & & & & & & \\ \hline
\hline
E4-D2-U1P &
\lw{0.020} & \lw{0.99} & \lw{$-0.0287$} &
\lw{$-2.23$} & \lw{$2.02$} &
\lw{$4.96\times10^{-6}$} & \lw{$4.04\times10^{-4}$}\\ 
(57.7, 0.91) & & & & & & & \\ \hline
E4-D2-U2P &
\lw{0.004} & \lw{0.85} & \lw{$-0.128$} &
\lw{$-2.23$} & \lw{$5.02$} &
\lw{$4.96\times10^{-6}$} & \lw{$2.51\times10^{-3}$}\\ 
(60.8, 0.53) & & & & & & & \\ \hline
E4-D2-U3P &
\lw{0.020} & \lw{0.99} & \lw{$-0.0133$} &
\lw{$-2.23$} & \lw{$2.17$} &
\lw{$4.97\times10^{-6}$} & \lw{$4.65\times10^{-4}$}\\ 
(58.1, 0.865) & & & & & & & \\ \hline
E4-D2-U4P &
\lw{0.003} & \lw{0.85} & \lw{$-0.109$} &
\lw{$-2.23$} & \lw{$5.30$} &
\lw{$4.97\times10^{-6}$} & \lw{$2.81\times10^{-3}$}\\ 
(60.7, 0.50) & & & & & & & \\ \hline
\hline
E4-D4-U1P &
\lw{0.014} & \lw{0.96} & \lw{$-1.09$} &
\lw{$-2.49$} & \lw{$2.09$} &
\lw{$5.01\times10^{-6}$} & \lw{$4.29\times10^{-4}$}\\ 
(55.3, 0.76) & & & & & & & \\ \hline
E4-D4-U2P &
\lw{0.019} & \lw{0.89} & \lw{$-1.49$} &
\lw{$-2.37$} & \lw{$4.16$} &
\lw{$3.39\times10^{-6}$} & \lw{$1.72\times10^{-3}$}\\ 
(62.0, 0.68) & & & & & & & \\ \hline
E4-D4-U3P &
\lw{0.020} & \lw{0.97} & \lw{$-1.08$} &
\lw{$-2.49$} & \lw{$2.24$} &
\lw{$5.01\times10^{-6}$} & \lw{$4.96\times10^{-4}$}\\ 
(62.6, 0.94) & & & & & & & \\ \hline
E4-D4-U4P &
\lw{0.018} & \lw{0.89} & \lw{$-1.45$} &
\lw{$-2.34$} & \lw{$4.26$} &
\lw{$3.36\times10^{-6}$} & \lw{$1.81\times10^{-3}$}\\ 
(56.1, 0.555) & & & & & & & \\ \hline
\hline
\end{tabular}

\newpage
\normalsize
\noindent
Table 2a: The ranges of parameters $r$ and $R$ to predict the 
desired neutrino mixings and masses   
for $m_bm_\tau>0$. 
\\
\vskip 5mm

\scriptsize
E1-D1-UiM (i=1,2,3,4)

\begin{tabular}{c|c||c||c|c|c}
\hline \hline
$ e , \mu$ & $ d , s$ & 
$ u , c , t $ & $r$ & $R$ $(10^{15})$ $(r)$ &
$\sin^2{2\theta_{\mu\tau}}$\\ 
\hline \hline
$(-,-)$&$(-,-)$&$(-,-),-$&
510-520, 550, 830-840, 900-950&0.55-0.99 (550)&0.92\\ 
&&$(-,+),-$&
510-520, 540, 830-840, 900-950&1.30-2.36 (830)&0.92 \\
&&$(+,-),-$&
510-520, 550, 830-840, 900-950&0.55-0.99 (550)&0.92\\
&&$(+,+),-$&
510-520, 540, 830-840, 900-950&1.30-2.36 (830)&0.92\\
\hline \hline
\end{tabular}

\vspace{10mm}
E1-D3-UiM (i=1,2,3,4)

\begin{tabular}{c|c||c||c|c|c}
\hline \hline
$ e , \mu$ & $ d , s$ & 
$ u , c , t $ & $r$ & $R$ $(10^{15})$ $(r)$&
$\sin^2{2\theta_{\mu\tau}}$\\ \hline \hline
$(-,-)$&$(+,-)$&$(-,-),-$&
550-580, 640-740, 840-910&0.84-1.51 (690)&0.91\\
&&$(-,+),-$&
560-580, 640-740, 840-900&0.87-1.57 (680)&0.92\\
&&$(+,-),-$&
550-580, 640-740, 840-910&0.84-1.51 (690)&0.91\\
&&$(+,+),-$&
560-580, 640-740, 840-900&0.87-1.57 (680)&0.92\\
\hline \hline
\end{tabular}

\vspace{10mm}
E3-D1-UiM (i=1,2,3,4)

\begin{tabular}{c|c||c||c|c|c}
\hline \hline
$ e , \mu$ & $ d , s$ & 
$ u , c , t $ & $r$ & $R$ $(10^{15})$ $(r)$&
$\sin^2{2\theta_{\mu\tau}}$\\ \hline \hline
$(+,-)$&$(-,-)$&$(-,-),-$&
510, 540, 840-850, 910-960&1.36-2.47 (840)&0.91\\
&&$(-,+),-$&
510, 540, 840-850, 910-950&1.38-2.51 (840)&0.92\\
&&$(+,-),-$&
510, 540, 840-850, 910-960&1.36-2.47 (840)&0.91\\
&&$(+,+),-$&
510, 540, 840-850, 910-950&1.38-2.51 (840)&0.92\\
\hline \hline
\end{tabular}

\vspace{10mm}
E3-D3-UiM (i=1,2,3,4)

\begin{tabular}{c|c||c||c|c|c}
\hline \hline
$ e , \mu$ & $ d , s$ & 
$ u , c , t $ & $r$ & $R$ $(10^{15})$ $(r)$&
$\sin^2{2\theta_{\mu\tau}}$\\  \hline \hline
$(+,-)$&$(+,-)$&$(-,-),-$&
520-540, 570-580, 790-810, 880-940&1.12-2.04 (790)&0.92\\
&&$(-,+),-$&
530, 570-580, 790-810, 880-930&0.63-1.13 (580)&0.93\\
&&$(+,-),-$&
520-540, 570-580, 790-810, 880-940&1.12-2.04 (790)&0.92\\
&&$(+,+),-$&
530, 570-580, 790-810, 880-930&0.63-1.13 (580)&0.93\\
\hline \hline
\end{tabular}

\vspace{10mm}
E1-D2-U2P
\vspace{1mm}

\begin{tabular}{c|c||c||c|c|c}
\hline \hline
$ e , \mu$ & $ d , s$ & 
$ u , c , t $ & $r$ & $R$ $(10^{13})$ $(r)$&
$\sin^2{2\theta_{\mu\tau}}$\\  \hline \hline
$(-,-)$&$(-,+)$&$(-,+),+$&
36.2-36.5&7.74-7.96 (36.5)&0.75\\
\hline \hline
\end{tabular} 
 
%%%%%%%%%%%%%%%%%%%%%%%%%%%%%%%%%%%%%%%%%%%%%%%%%%%%%%%%%%%%%%%%%%%
\newpage
\normalsize
\noindent
Table 2b: Predictions of neutrino mixing angles and masses 
for $m_bm_\tau>0$. All cases correspond to Table 2a. 

\scriptsize
\begin{flushleft}
 \begin{tabular}{c|c|c|c|c|c|c|c}
\hline \hline
Model &
\lw{$\sin^2{2\theta_{e\mu}}$}&     %% atmospheric
\lw{$\sin^2{2\theta_{\mu\tau}}$} & %% solar (MSW small) 
\lw{$m_1$($10^{-3}$eV)}&           %% m(nu_e)
\lw{$m_2$($10^{-3}$eV)}&           %% m(nu_mu)
\lw{$m_3$($10^{-2}$eV)}&           %% m(nu_tau) 
\lw{$\Delta m_{12}^2$ (eV$^2$)}&
\lw{$\Delta m_{23}^2$ (eV$^2$)} \\
$(r , R/10^{15})$ & & & & & & & \\ \hline \hline
E1-D1-U1M &
\lw{0.018} & \lw{0.92} & \lw{$0.976$} &
\lw{$2.44$} & \lw{$-2.63$} &
\lw{$4.99\times10^{-6}$} & \lw{$6.90\times10^{-4}$}\\ 
(550, 0.77) & & & & & & & \\ \hline
E1-D1-U2M &
\lw{0.016} & \lw{0.92} & \lw{$0.941$} &
\lw{$2.43$} & \lw{$-2.45$} &
\lw{$5.01\times10^{-6}$} & \lw{$5.96\times10^{-4}$}\\ 
(830, 1.83) & & & & & & & \\ \hline
E1-D1-U3M &
\lw{0.019} & \lw{0.92} & \lw{$0.979$} &
\lw{$2.44$} & \lw{$-2.64$} &
\lw{$4.99\times10^{-6}$} & \lw{$6.90\times10^{-4}$}\\ 
(550, 0.77) & & & & & & & \\ \hline
E1-D1-U4M &
\lw{0.016} & \lw{0.92} & \lw{$0.942$} &
\lw{$2.43$} & \lw{$-2.45$} &
\lw{$5.01\times10^{-6}$} & \lw{$5.96\times10^{-4}$}\\ 
(830, 1.83) & & & & & & & \\ \hline 
\hline
E1-D3-U1M &
\lw{0.009} & \lw{0.91} & \lw{$2.30\times10^{-4}$} &
\lw{$2.24$} & \lw{$-2.72$} &
\lw{$5.00\times10^{-6}$} & \lw{$7.33\times10^{-4}$}\\ 
(690, 1.175) & & & & & & & \\ \hline
E1-D3-U2M &
\lw{0.009} & \lw{0.92} & \lw{$2.34\times10^{-4}$} &
\lw{$2.23$} & \lw{$-2.49$} &
\lw{$4.99\times10^{-6}$} & \lw{$6.15\times10^{-4}$}\\ 
(680, 1.22) & & & & & & & \\ \hline
E1-D3-U3M &
\lw{0.010} & \lw{0.91} & \lw{$2.49\times10^{-4}$} &
\lw{$2.24$} & \lw{$-2.72$} &
\lw{$5.01\times10^{-6}$} & \lw{$7.33\times10^{-4}$}\\ 
(690, 1.175) & & & & & & & \\ \hline
E1-D3-U4M &
\lw{0.009} & \lw{0.92} & \lw{$2.21\times10^{-4}$} &
\lw{$2.24$} & \lw{$-2.49$} &
\lw{$5.00\times10^{-6}$} & \lw{$6.15\times10^{-4}$}\\ 
(680, 1.22) & & & & & & & \\ \hline 
\hline
E3-D1-U1M &
\lw{0.015} & \lw{0.91} & \lw{$1.16$} &
\lw{$2.52$} & \lw{$-2.44$} &
\lw{$5.00\times10^{-6}$} & \lw{$5.91\times10^{-4}$}\\ 
(840, 1.915) & & & & & & & \\ \hline
E3-D1-U2M &
\lw{0.017} & \lw{0.92} & \lw{$1.14$} &
\lw{$2.51$} & \lw{$-2.36$} &
\lw{$5.01\times10^{-6}$} & \lw{$5.50\times10^{-4}$}\\ 
(840, 1.945) & & & & & & & \\ \hline
E3-D1-U3M &
\lw{0.015} & \lw{0.91} & \lw{$1.16$} &
\lw{$2.52$} & \lw{$-2.44$} &
\lw{$5.00\times10^{-6}$} & \lw{$5.91\times10^{-4}$}\\ 
(840, 1.915) & & & & & & & \\ \hline
E3-D1-U4M &
\lw{0.017} & \lw{0.92} & \lw{$1.14$} &
\lw{$2.51$} & \lw{$-2.36$} &
\lw{$5.00\times10^{-6}$} & \lw{$5.50\times10^{-4}$}\\ 
(840, 1.945) & & & & & & & \\ \hline 
\hline
E3-D3-U1M &
\lw{0.016} & \lw{0.92} & \lw{$0.241$} &
\lw{$2.25$} & \lw{$-2.65$} &
\lw{$5.00\times10^{-6}$} & \lw{$6.96\times10^{-4}$}\\ 
(790, 1.58) & & & & & & & \\ \hline
E3-D3-U2M &
\lw{0.016} & \lw{0.93} & \lw{$0.232$} &
\lw{$2.25$} & \lw{$-2.50$} &
\lw{$4.99\times10^{-6}$} & \lw{$6.19\times10^{-4}$}\\ 
(580, 0.88) & & & & & & & \\ \hline
E3-D3-U3M &
\lw{0.016} & \lw{0.92} & \lw{$0.242$} &
\lw{$2.25$} & \lw{$-2.65$} &
\lw{$5.00\times10^{-6}$} & \lw{$6.96\times10^{-4}$}\\ 
(790, 1.58) & & & & & & & \\ \hline
E3-D3-U4M &
\lw{0.017} & \lw{0.93} & \lw{$0.233$} &
\lw{$2.25$} & \lw{$-2.50$} &
\lw{$5.00\times10^{-6}$} & \lw{$6.19\times10^{-4}$}\\ 
(580, 0.88) & & & & & & & \\ \hline
\hline
\end{tabular}
\end{flushleft}
\vspace{5mm}

\begin{flushleft}
 \begin{tabular}{c|c|c|c|c|c|c|c}
\hline \hline
Model &
\lw{$\sin^2{2\theta_{e\mu}}$}&     %% atmospheric
\lw{$\sin^2{2\theta_{\mu\tau}}$} & %% solar (MSW small) 
\lw{$m_1$($10^{-3}$eV)}&           %% m(nu_e)
\lw{$m_2$($10^{-3}$eV)}&           %% m(nu_mu)
\lw{$m_3$($10^{-2}$eV)}&           %% m(nu_tau) 
\lw{$\Delta m_{12}^2$ (eV$^2$)}&
\lw{$\Delta m_{23}^2$ (eV$^2$)} \\
$(r , R/10^{15})$ & & & & & & & \\ \hline \hline
E1-D2-U2P &
\lw{0.006} & \lw{0.75} & \lw{$-9.47\times10^{-4}$} &
\lw{$3.12$} & \lw{$-1.47$} &
\lw{$9.71\times10^{-6}$} & \lw{$2.06\times10^{-4}$}\\ 
(36.5, 0.0785) & & & & & & & \\ \hline
\hline
\end{tabular}
\end{flushleft}

\newpage
\normalsize
\noindent
Figure captions:

\begin{itemize}
\item[Fig.1:] The parameter $r$ dependence of neutrino mixing angles and 
the ratio neutrino masses squared differences for E1-D3-U2P case 
($m_bm_\tau<0$). 
Fig.1a, 1b and 1c show 
the $r$ dependence of $\sin^2 2\theta_{\mu\tau}$, 
$\sin^2 2\theta_{e\mu}$ 
and $\Delta m_{12}^2/\Delta m_{23}^2$. The allowed region of $r$ is 
essentially determined by reproducing $\sin^2 2\theta_{e\mu}$. 

\item[Fig.2:] The same as Fig.1 for E4-D2-U3P case ($m_bm_\tau<0$). 
  The allowed region of $r$ is 
essentially determined by reproducing $\sin^2 2\theta_{e\mu}$. 

\item[Fig.3:]  The same as Fig.1 for E1-D3-U4M case ($m_bm_\tau>0$). 
  The allowed region of $r$ is 
essentially determined by reproducing $\sin^2 2\theta_{e\mu}$. 

\item[Fig.4:]  The same as Fig.1 for E3-D3-U3M case ($m_bm_\tau>0$). 
  The allowed region of $r$ is 
essentially determined by reproducing $\sin^2 2\theta_{e\mu}$. 
 
 \item[Fig.5:]  The same as Fig.1 for E1-D2-U2P case ($m_bm_\tau>0$). 
  The allowed region of $r$ is 
essentially determined by reproducing $\sin^2 2\theta_{\mu\tau}$ and 
$\Delta m_{12}^2/\Delta m_{23}^2$. The allowed region is a tiny one, 
so that we need a fine tuning of $r$.
\end{itemize}

\newpage

\vspace*{-2.5cm}
\centerline{\epsfxsize=9.5cm\epsfbox{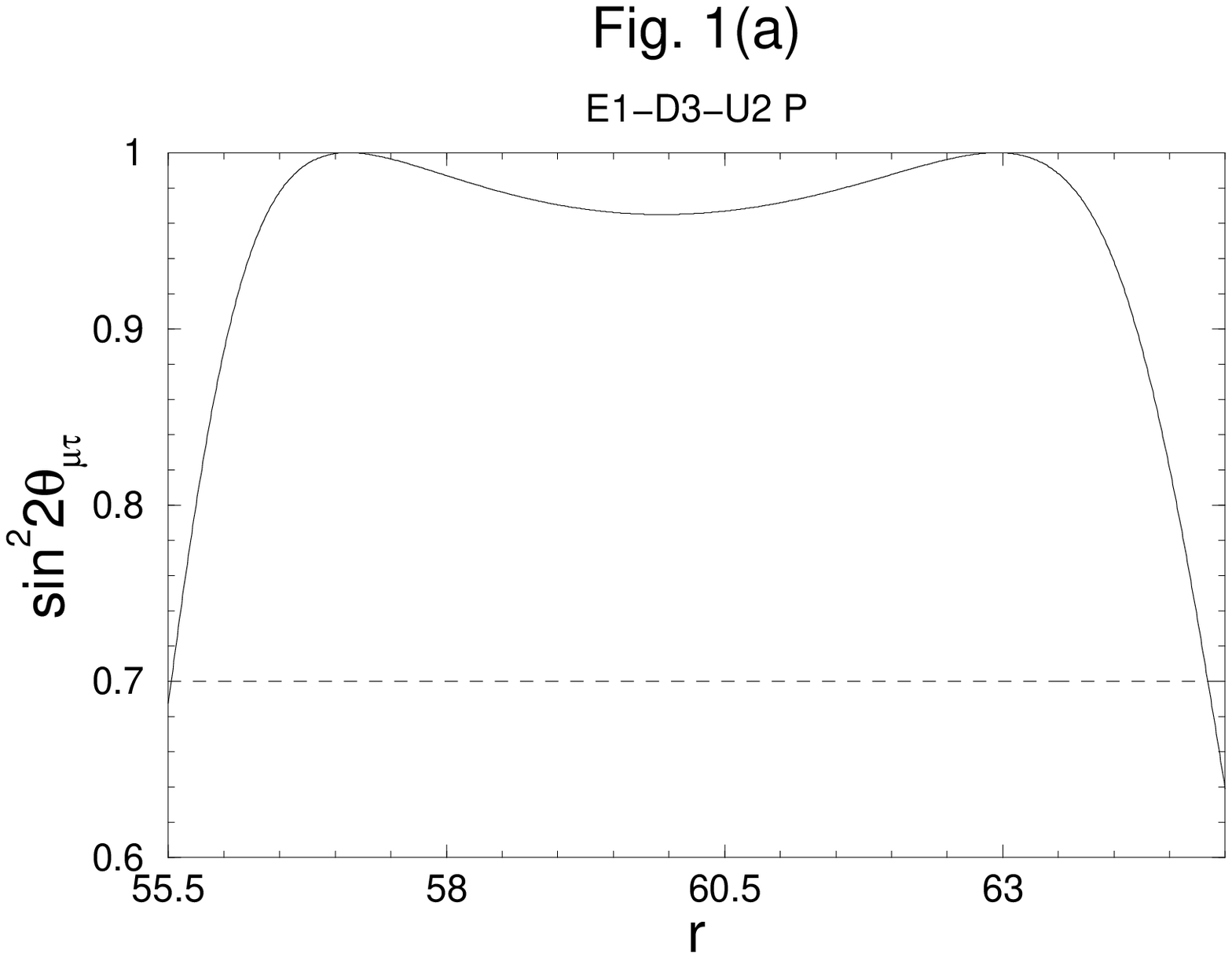}}
\medskip
\centerline{\epsfxsize=9.5cm\epsfbox{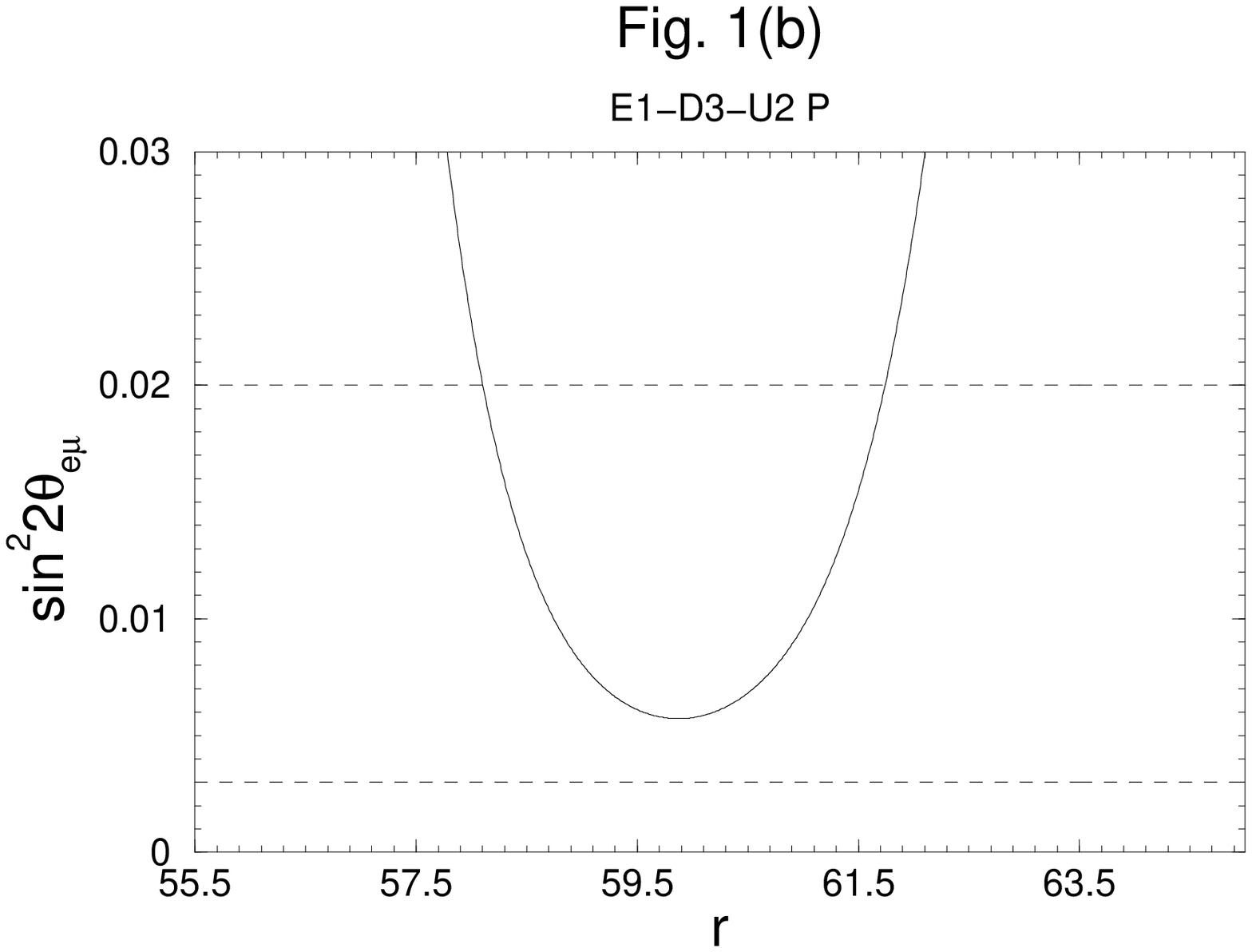}}
\medskip
\centerline{\epsfxsize=9.5cm\epsfbox{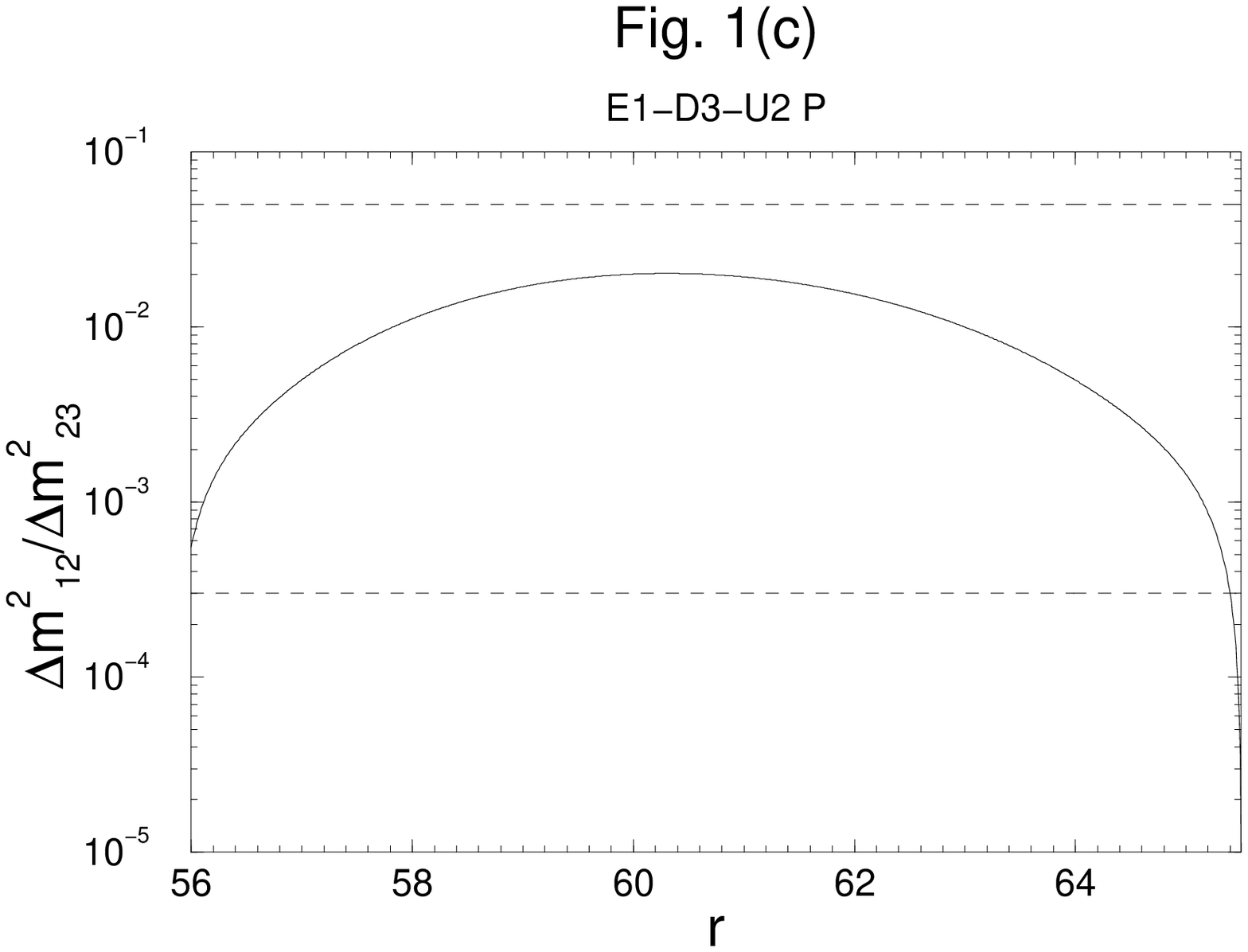}}

\newpage

\vspace*{-2.5cm}
\centerline{\epsfxsize=9.5cm\epsfbox{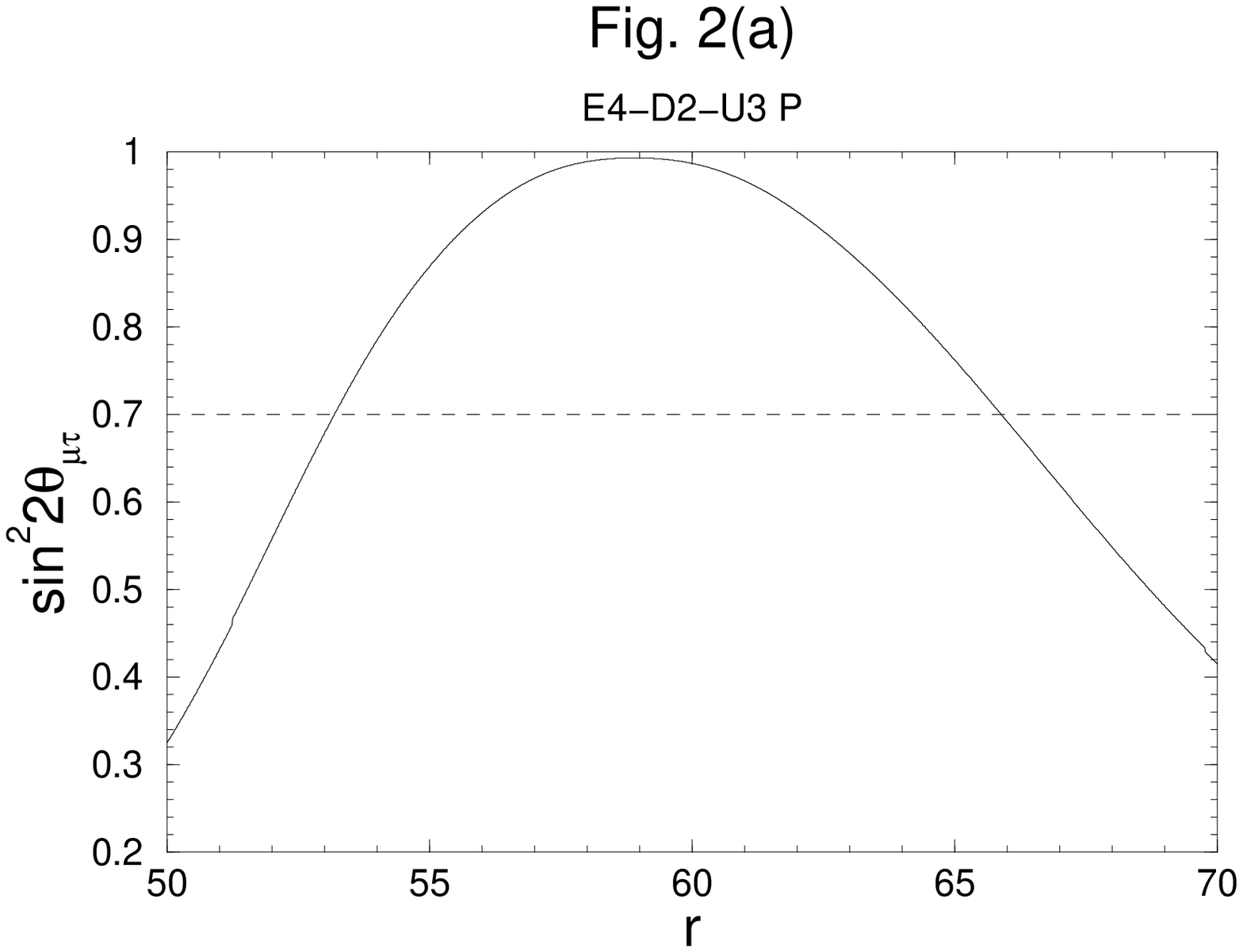}}
\medskip
\centerline{\epsfxsize=9.5cm\epsfbox{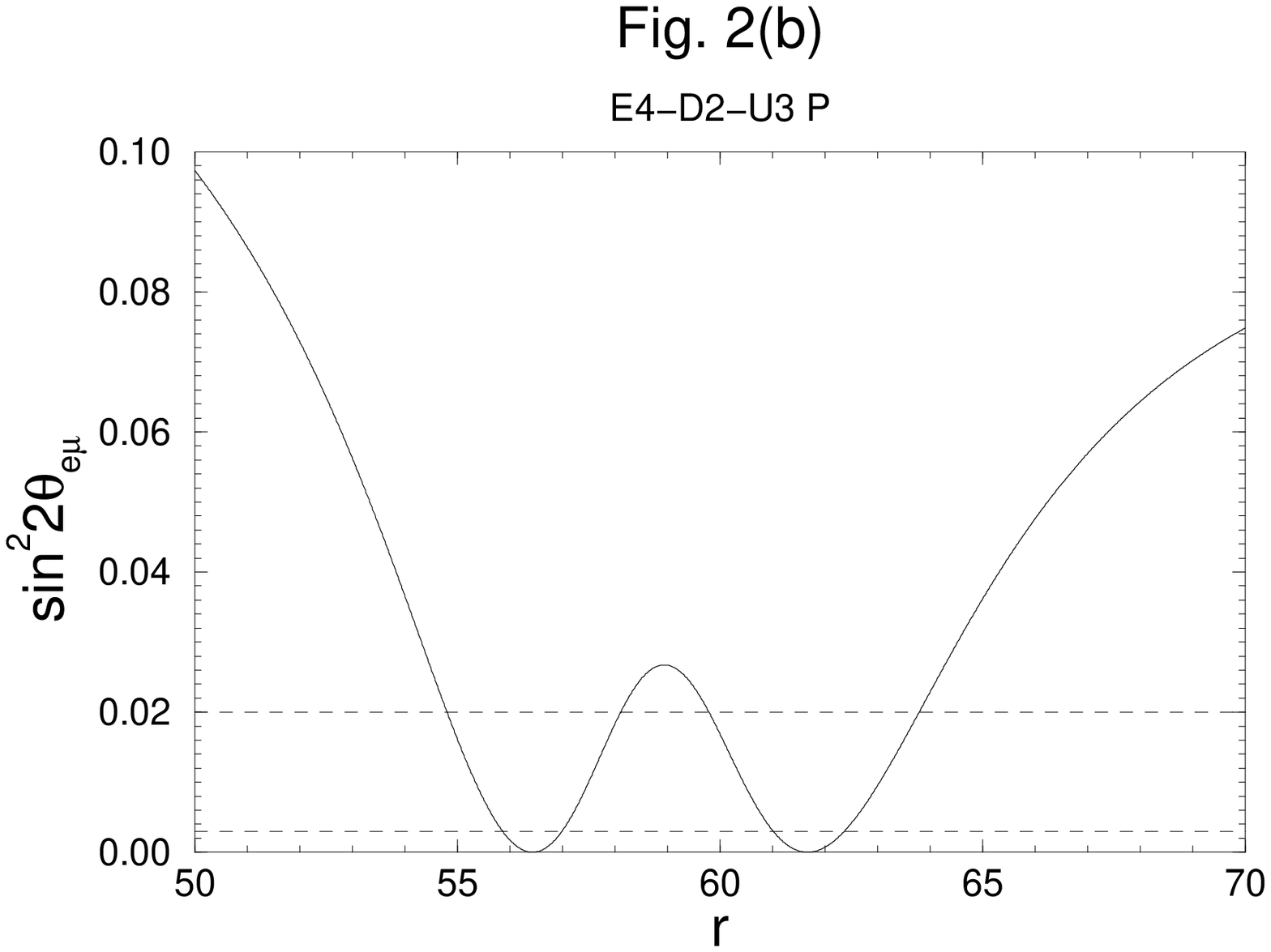}}
\medskip
\centerline{\epsfxsize=9.5cm\epsfbox{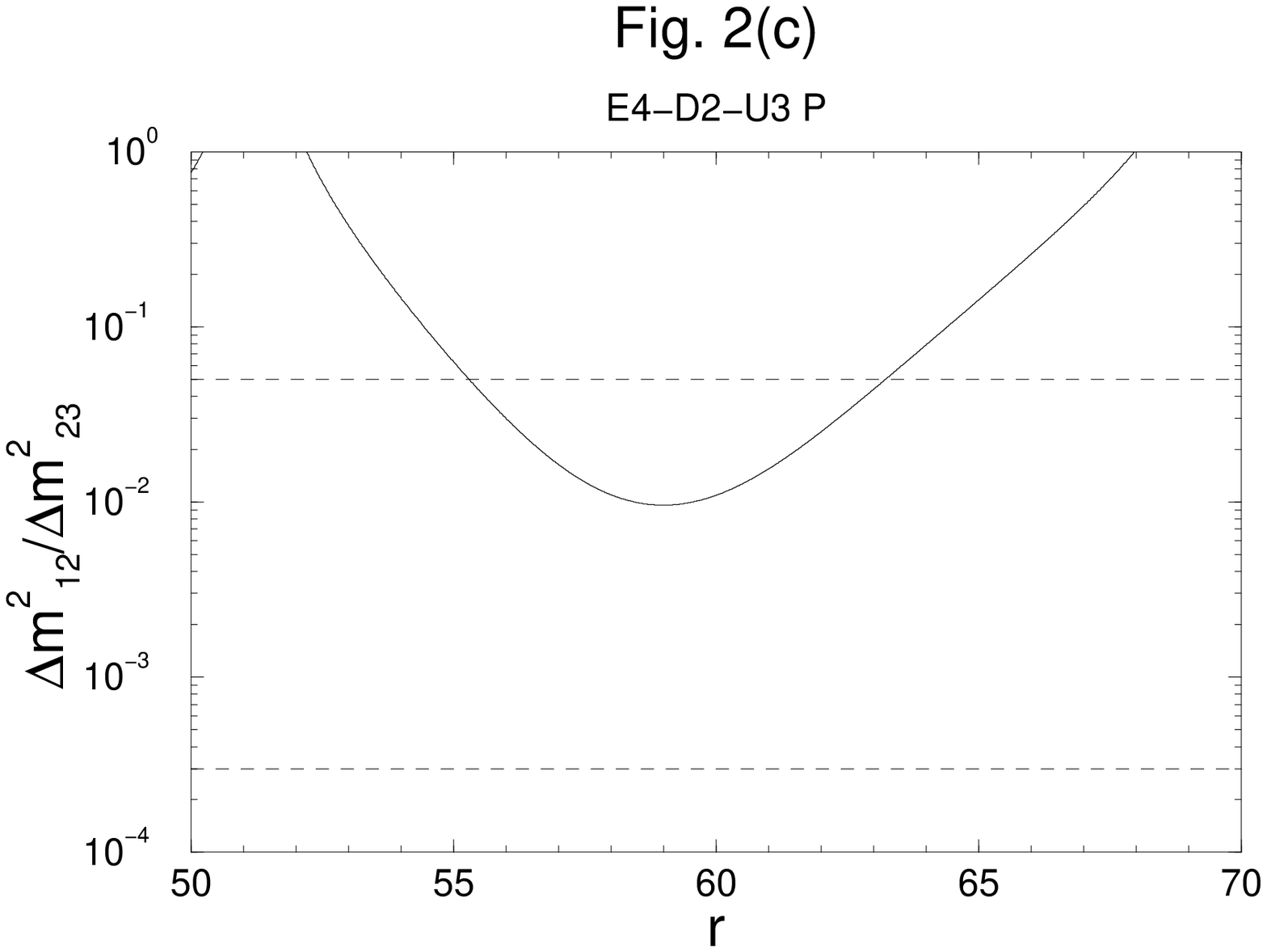}}

\newpage

\vspace*{-2.5cm}
\centerline{\epsfxsize=9.5cm\epsfbox{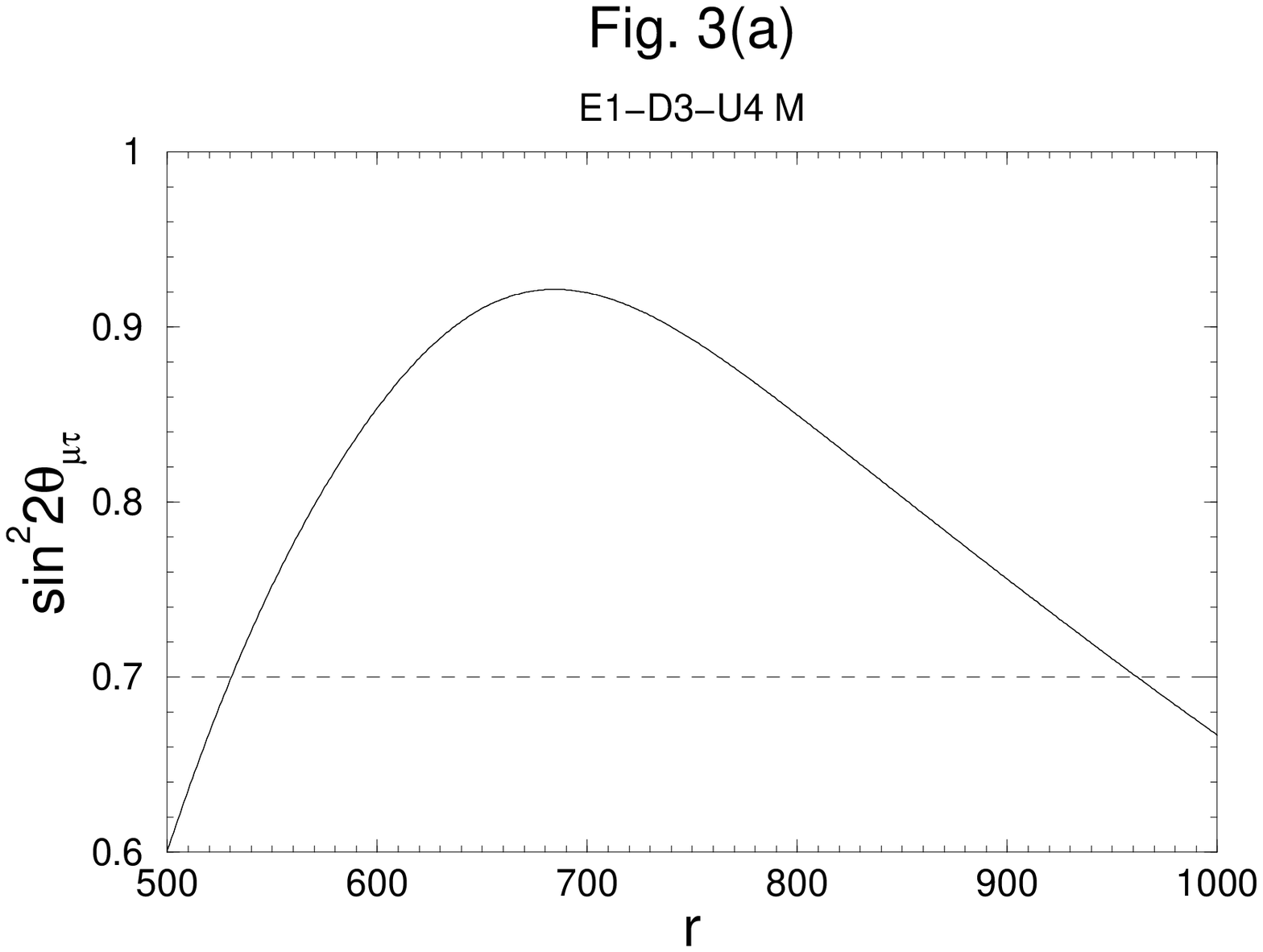}}
\medskip
\centerline{\epsfxsize=9.5cm\epsfbox{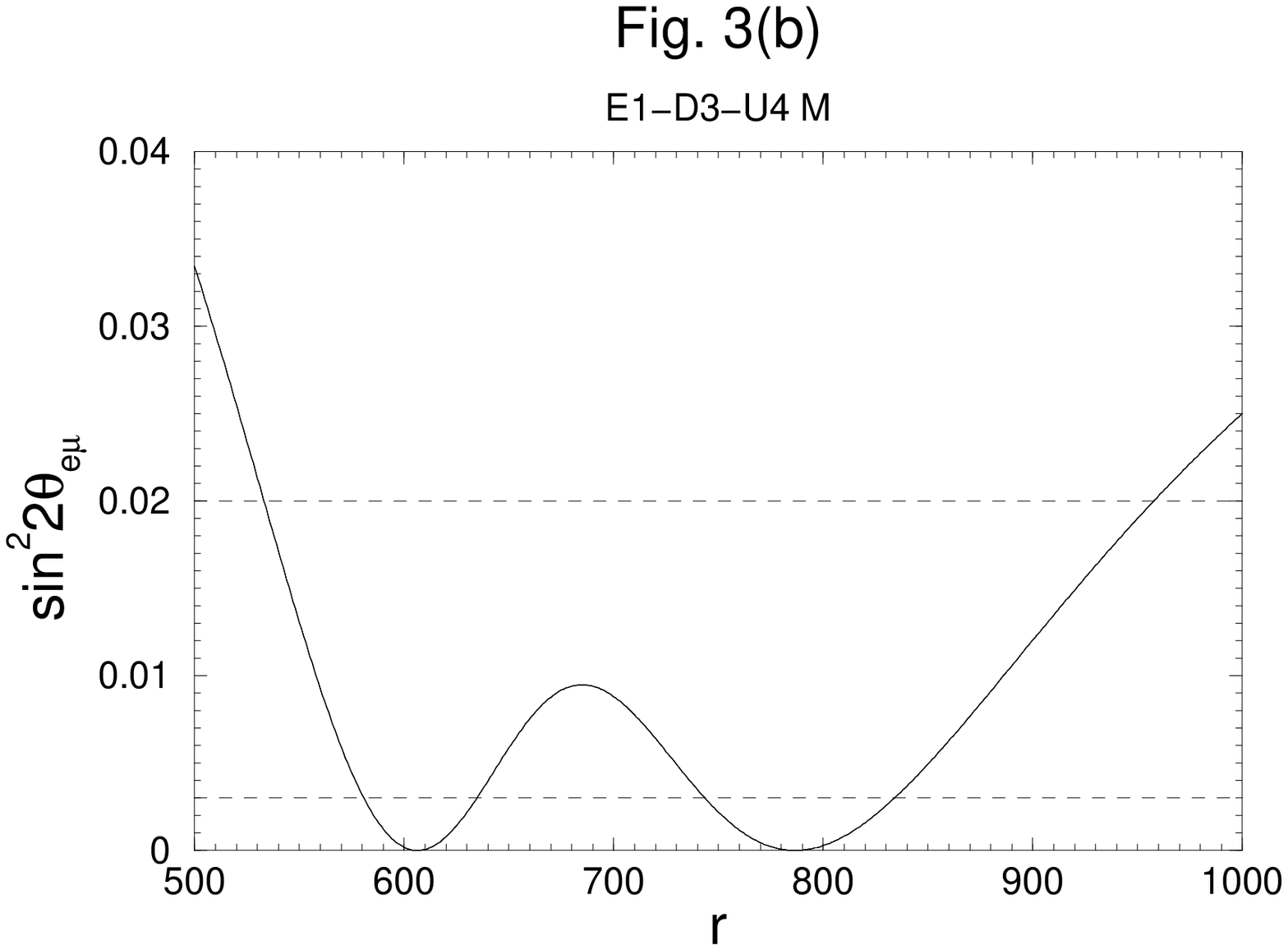}}
\medskip
\centerline{\epsfxsize=9.5cm\epsfbox{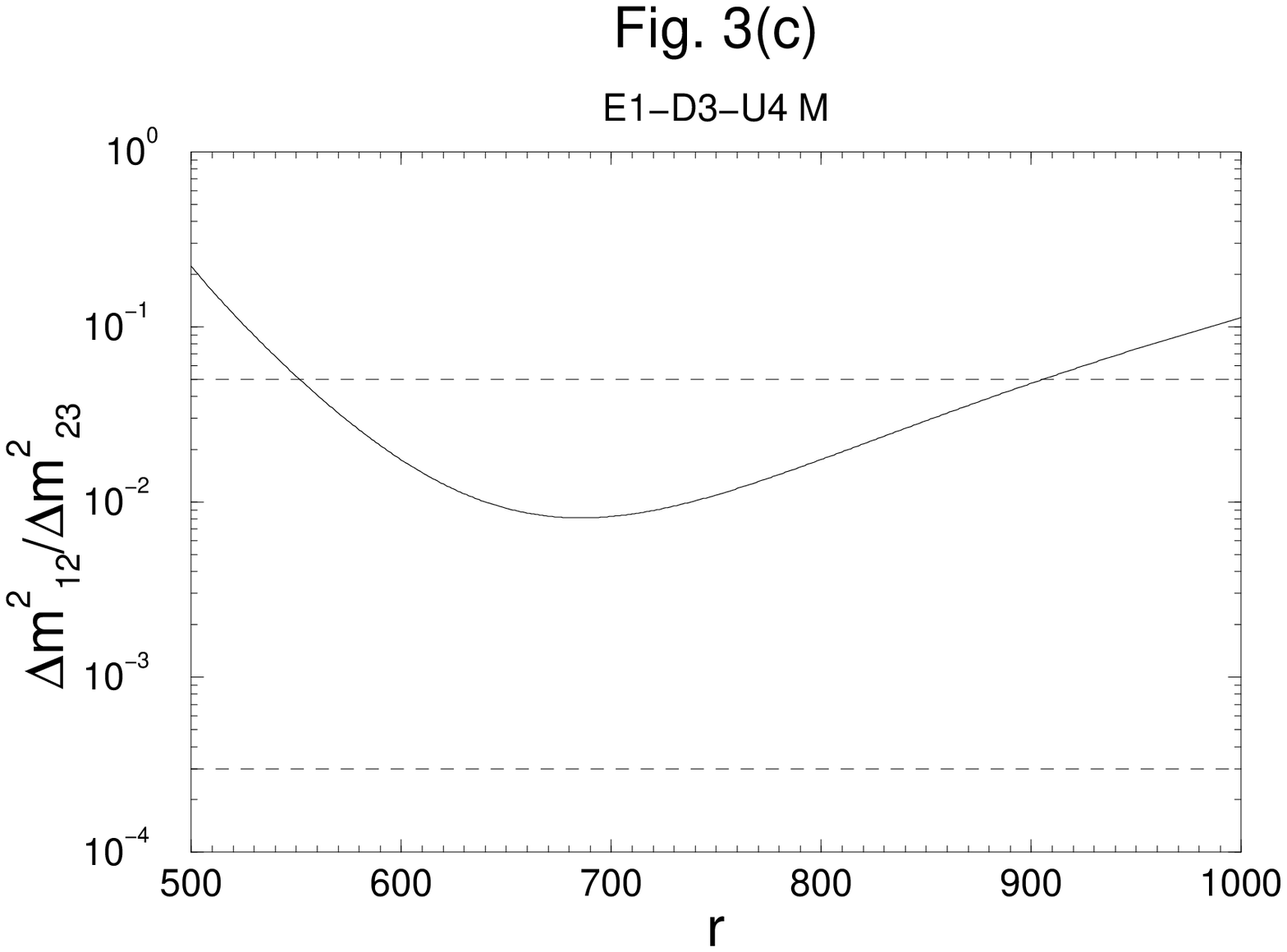}}

\newpage

\vspace*{-2.5cm}
\centerline{\epsfxsize=9.5cm\epsfbox{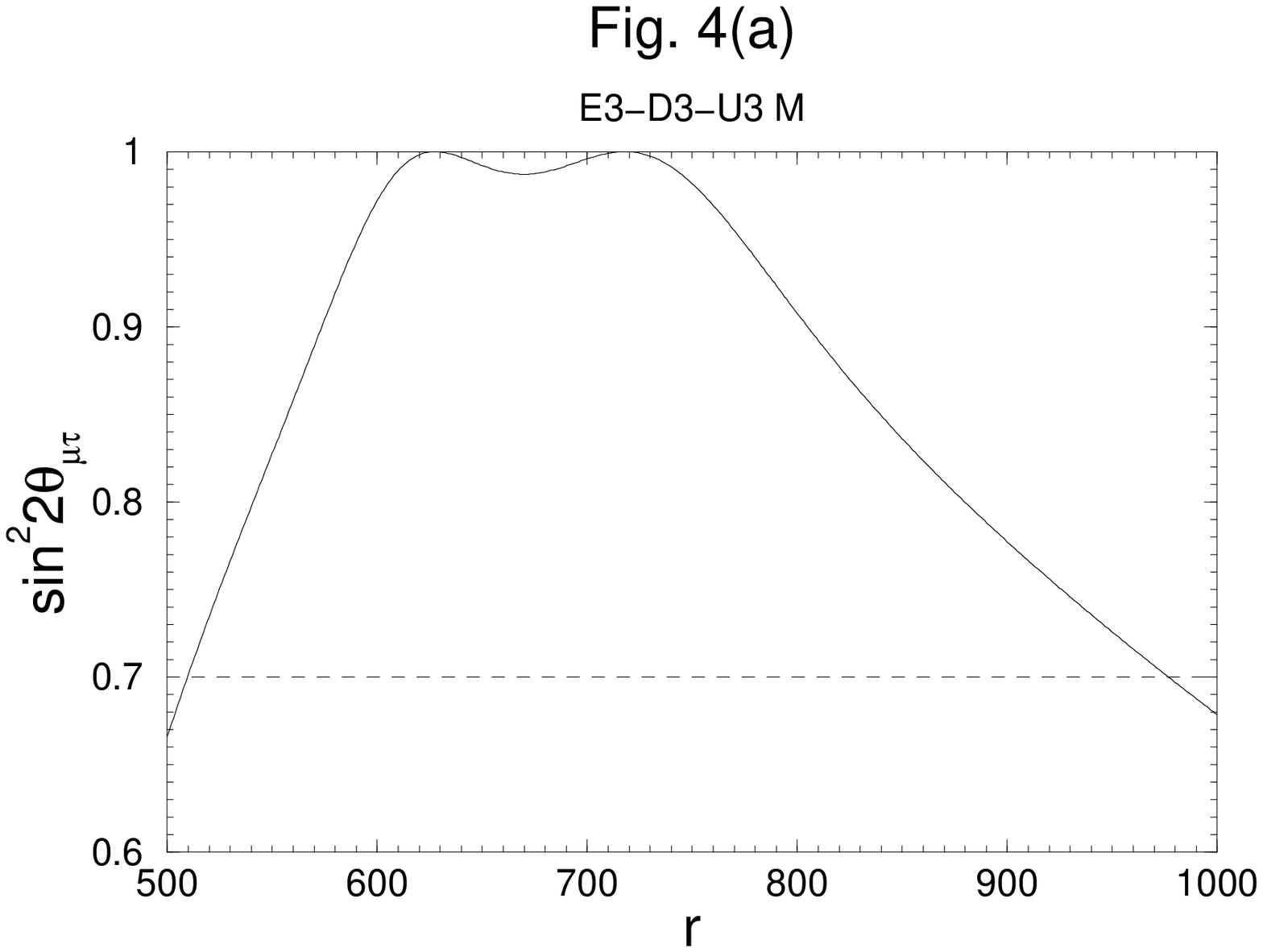}}
\medskip
\centerline{\epsfxsize=9.5cm\epsfbox{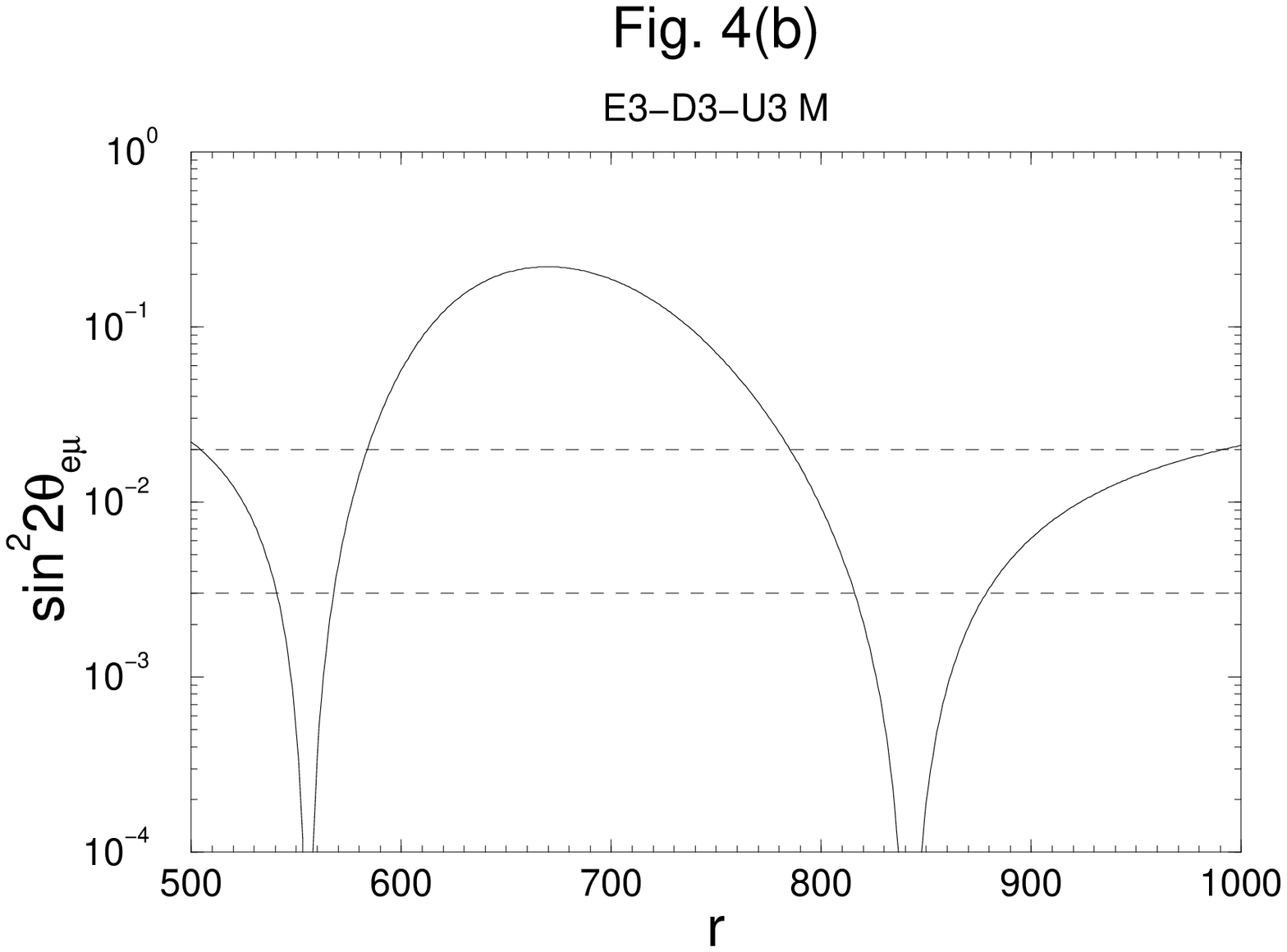}}
\medskip
\centerline{\epsfxsize=9.5cm\epsfbox{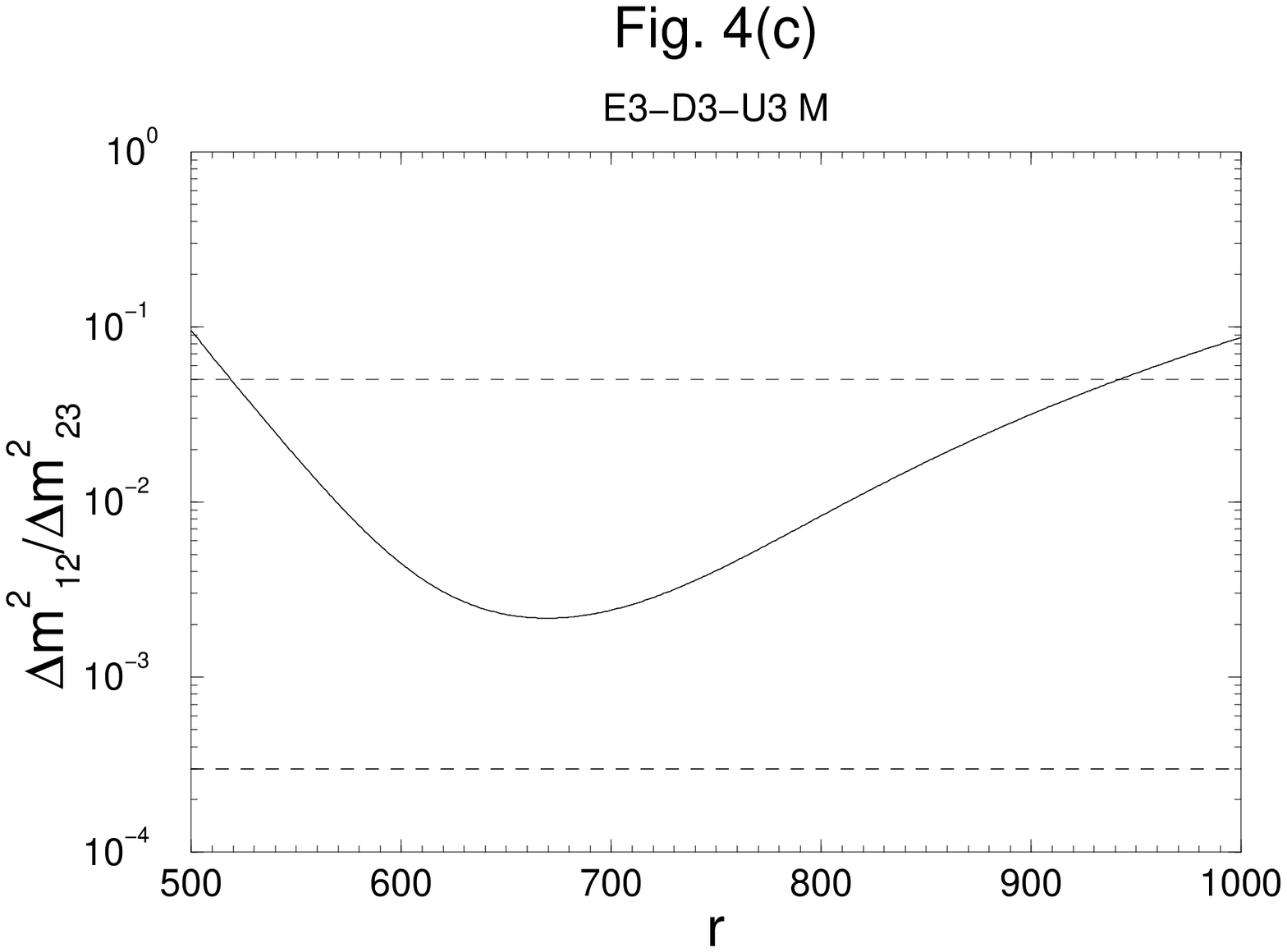}}

\newpage

\vspace*{-2.5cm}
\centerline{\epsfxsize=9.5cm\epsfbox{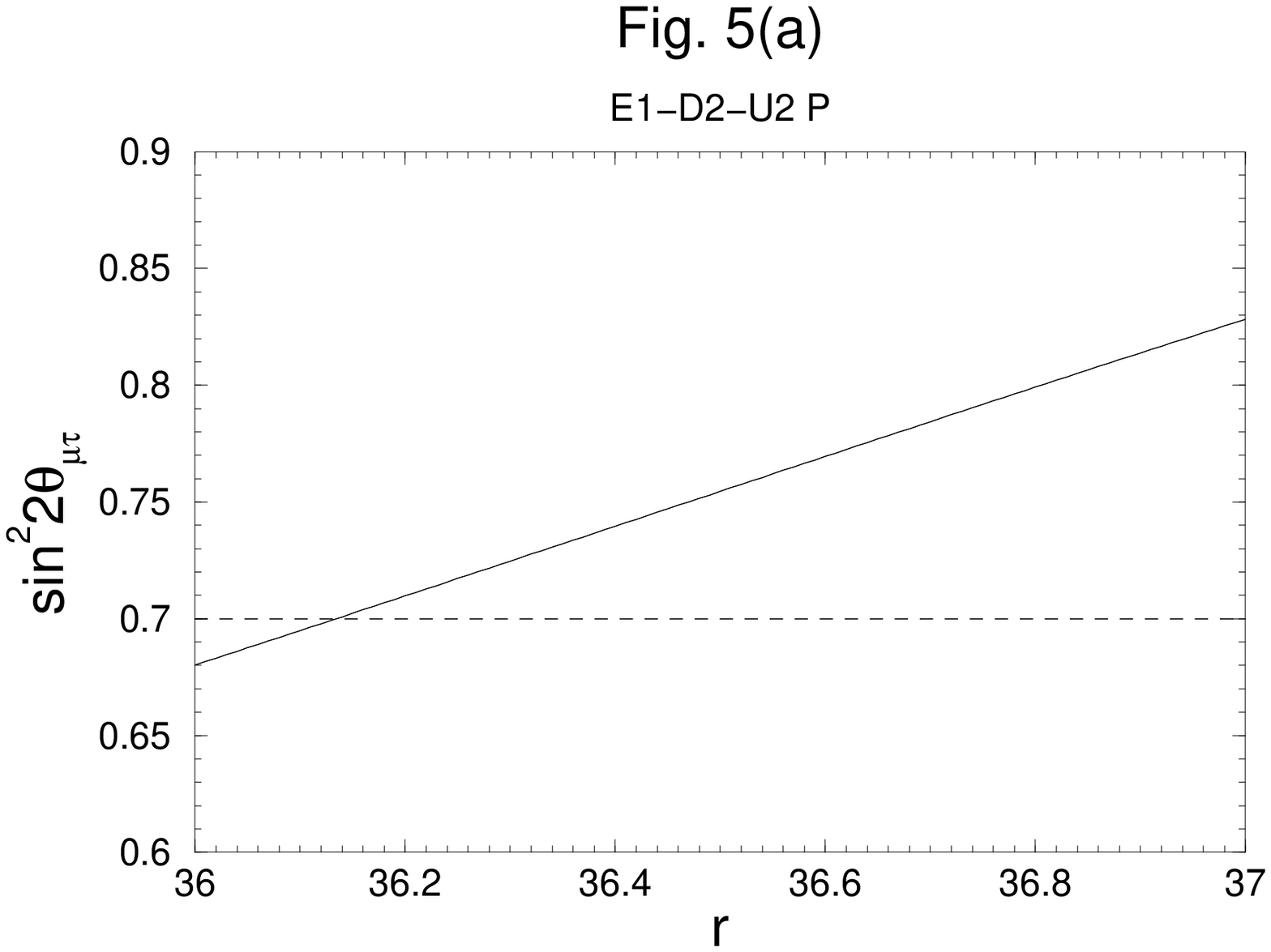}}
\medskip
\centerline{\epsfxsize=9.5cm\epsfbox{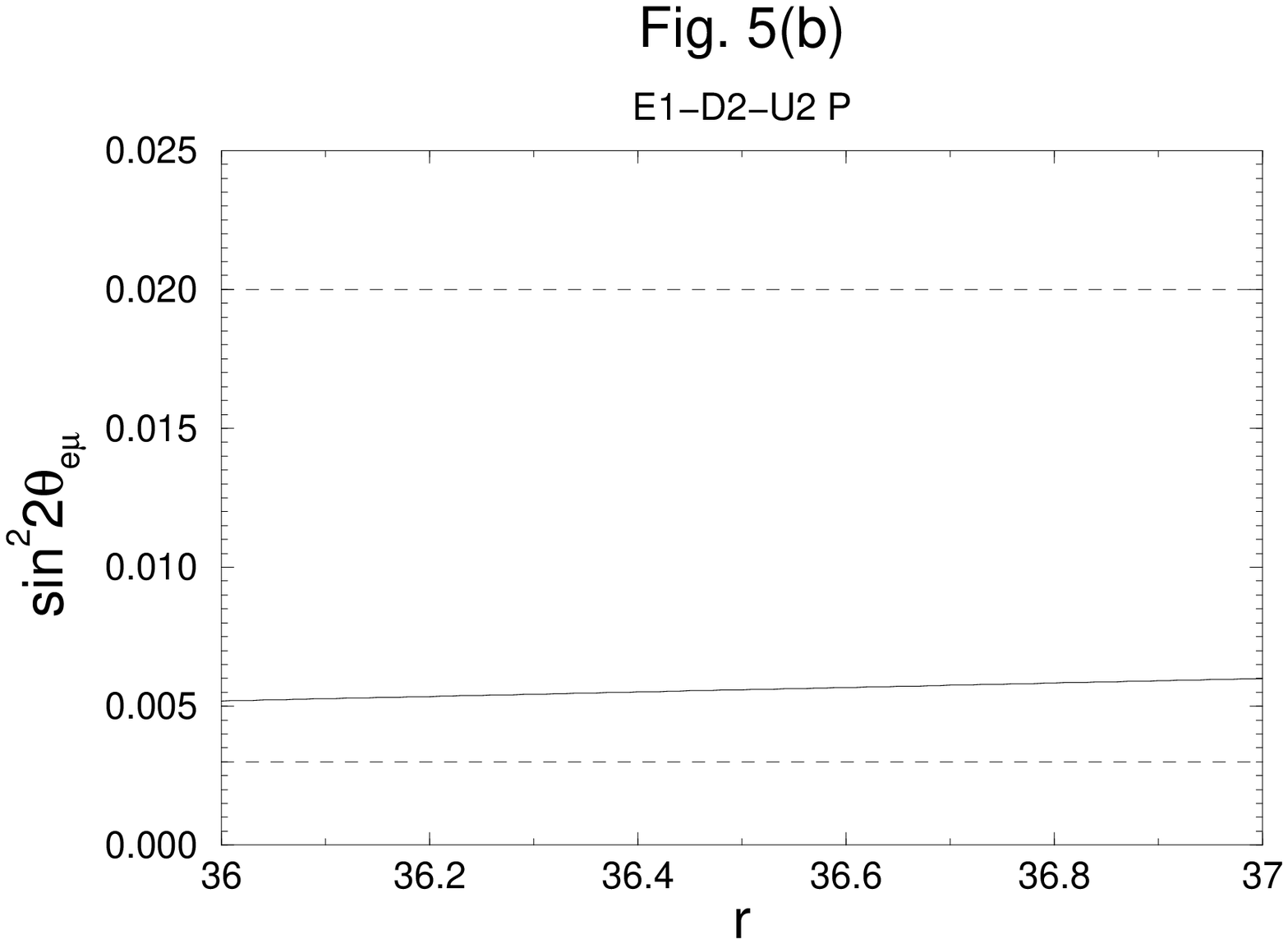}}
\medskip
\centerline{\epsfxsize=9.5cm\epsfbox{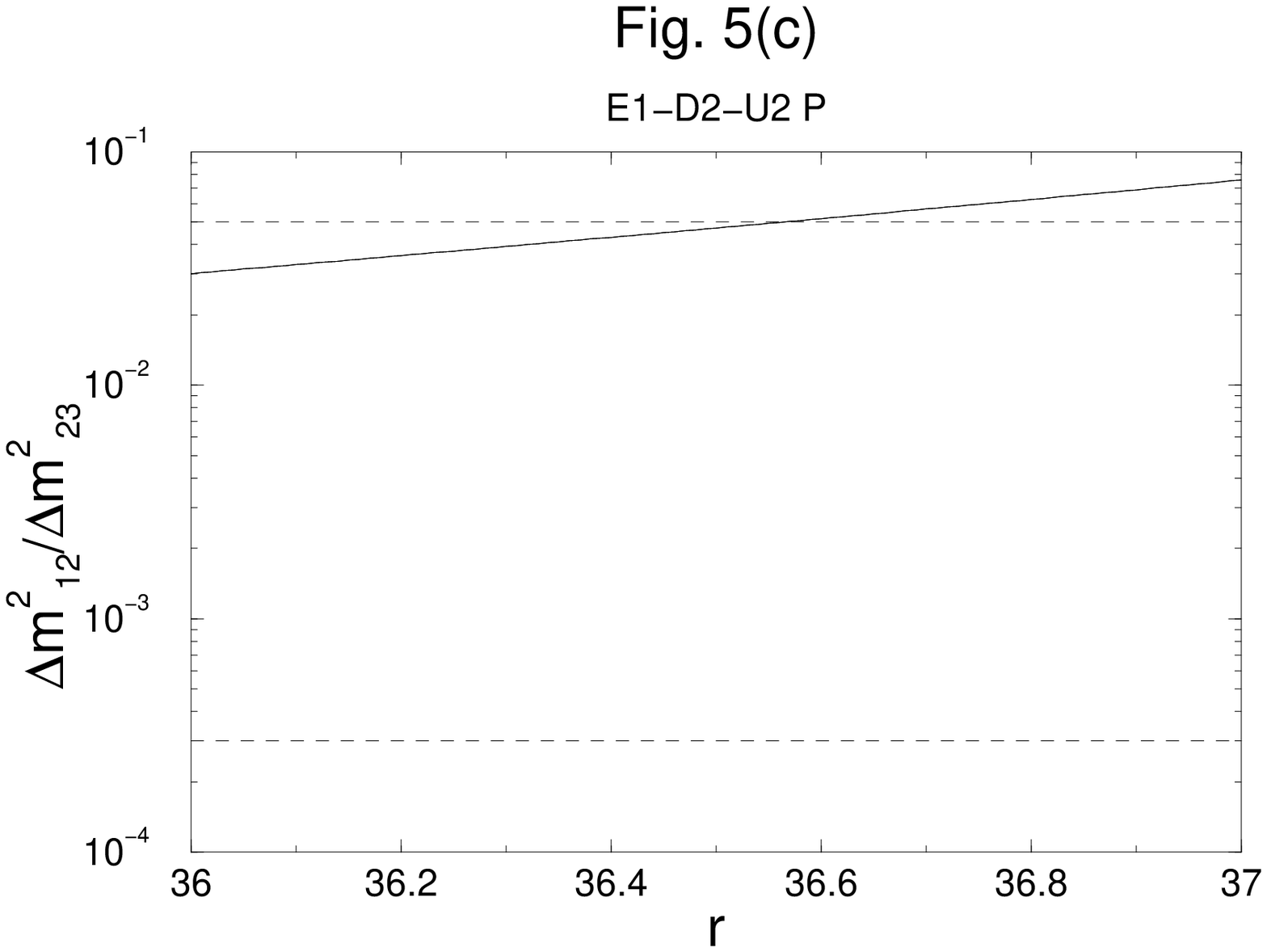}}


\newpage

\begin{thebibliography}{99}
\bibitem{Kajita}T. Kajita, a talk given at Neutrino 98, 
June 4-9, 1998 Takayama; hep-ex/9807003. 
\bibitem{Bemporad}C. Bemporad,  a talk given at Neutrino 98, 
June 4-9, 1998 Takayama; \PL{B420}, 397 (1998). 
\bibitem{Suzuki}Y. Suzuki, a talk given at Neutrino 98, 
June 4-9, 1998 Takayama.  
\bibitem{BMoh}K.S. Babu and R. N. Mohapatra, \PRL{70}, 2845 (1993). 
\bibitem{LMoh}D-G. Lee and R. N. Mohapatra, \PR{D51}, 1353 (1995). 
\bibitem{BShafi}K.S. Babu and Q. Shafi, \PL{B294}, 235 (1992). 
\bibitem{AG}Y. Achiman and T. Greiner, \PL{B329}, 33 (1994);\\
\NP{B443}, 3 (1995).
\bibitem{BM}B. Brahmachari and R. N. Mohapatra, \PR{D58}, 015003 (1998).
\bibitem{FK}H. Fusaoka and Y. Koide, \PR{D57}, 3986 (1998).
\bibitem{HL}N. Hata and P. Langacker, \PR{56}, 6107 (1997). 
\end{thebibliography}
\end{document}